%% file: survey-paper.tex
\documentclass[journal]{IEEEtran}

\usepackage{myArticleStyle}
\usepackage{makecell}
\usepackage{longtable}
\begin{document}
		
	\title{Adaptive Cruise Control in Autonomous Vehicles: Challenges, Gaps, Comprehensive Review, and, Future Directions}
	
	\input{includes/sections/authors}	
	\maketitle 
	
	\input{includes/sections/abstract}
	\input{includes/sections/keywords}
	
	\input{includes/sections/introduction}

\input{includes/sections/background}

	\input{includes/sections/literatureSurvey}	
	\input{includes/sections/discussions}
	\input{includes/sections/future}	
	\input{includes/sections/concluslions}
	\input{includes/sections/acknowledgement}
	
	\bibliographystyle{IEEEtran}
	\bibliography{paper}		
\end{document}

%% file: includes/sections/authors.tex
\author
{
	Shradha Bavalatti, 
	Yash Kangralkar, 
	Santosh Pattar, and
	Veena P Badiger
		
	\thanks
	{
		All authors are with the Department of Computer Science and Engineering, KLE Technological University, Belagavi, India. (e-mail: shradhabavalatti@gmail.com, yashkangralkar2510@gmail.com santoshpattar01@gmail.com, and, veena.badiger92@gmail.com)
	}%
}

%% file: includes/sections/abstract.tex
\begin{abstract}

The development of Autonomous Vehicles (AVs) has redefined the way of transportation by eliminating the need for human intervention in driving. This revolution is fueled by rapid advancements in adaptive cruise control (ACC), which make AVs capable of interpreting their surroundings and responding intelligently. While AVs offer significant advantages, such as enhanced safety and improved traffic efficiency, they also face several challenges that need to be addressed. Existing survey papers often lack a comprehensive analysis of these challenges and their potential solutions. Our paper stands out by meticulously identifying these gaps in current ACC research and offering impactful future directions to guide researchers in designing next-generation ACC systems. Our survey provides a detailed and systematic review, addressing the limitations of previous studies and proposing innovative approaches to achieve sustainable and fault-resilient urban transportation.
\end{abstract}

%% file: includes/sections/keywords.tex

%% file: includes/sections/introduction.tex
\section{Introduction}
\label{sec:intro}
	\IEEEPARstart{I}{n} the recent years, transportation sector has shifted towards incorporating autonomous technologies for maintaining a safe and optimized ecosystem. Amongst
	several innovations, the emergence of autonomous vehicles (AVs) \cite{Autonomousvehicles}, Electric Vehicles (EVs), and hybrid vehicles have gained prominent focus. AVs have redefined the concept of transportation by providing self-driven capabilities that eliminate the need for human intervention. Meanwhile, the proliferation of EVs and hybrid vehicles has shown a significant step towards reducing carbon footprints \cite{ACCinEV}. As these transformative technologies continue to gain attention, it becomes vital to extend research on various functionalities of these technologies in order to move closer to achieving the goal of sustainability \cite{Autonomousvehicleimpacts}. 
	\par 
	The concept of automotive driving in AVs is achieved through the Advanced Driver Assistance System (ADAS) \cite{ADASINTRO}, it stands as the backbone technology stack of AVs. Its a set of integrated technologies that work together to uplift the vehicle's safety, and driving comfort to improve the overall performance of the vehicle. It utilizes a set of sensors, actuators, control units, and algorithms that analyze the vehicle's environment and execute appropriate actions. ADAS consists of a wide range of features. One of which is Adaptive Cruise Control (ACC) \cite{ACCINTRO}, a technology that enables autonomous vehicle navigation. It represents the evolution of traditional cruise control by adding features such as adjusting vehicle speed, avoiding collision, maintaining traffic flow \textit{etc.} These features ensure the driver a hassle-free and reliant driving experience and enhance road safety by reducing rear-end collisions caused by human errors. As ACC continues to evolve, it emerges as a prominent technology in AVs.
	\par 
    Although ACC offers a diverse range of advantages, it also has certain pitfalls like inaccurate sensor readings, complex traffic scenarios, maintaining string stability, V2V communication \textit{etc.}  \cite{ACC1} Therefore, these issues need to be addressed to maximize the benefit of ACC. \cite{Larsson2012Driver}  
	\par 
	Logically, in ACC two variables are involved in the simplest scenario. A vehicle of interest ("ego car") follows the "lead car". Ego car is embedded with sensors that capture the speed and distance of the lead car. Using this data, an ACC algorithm called as Model Predictive Controller (MPC) computes appropriate acceleration or deceleration for the ego car. The working and performance of MPC leads to smooth driving experience in different traffic conditions. \cite{ACCSIM}
    \par 
    In the past, several reviews have been conducted on the ADAS and ACC. Most of them concentrate on the string stability of the platoon \cite{Gunter2020Model-Based}, and few others address the security of the vehicle like (Alotibi \textit{et. al.} \cite{Alotibi2021Anomaly}) address the attacks targeting the ACC, (Zhang \textit{et. al.} \cite{Zhang2023Adaptive}). Whereas, the rest are distributed on fields like road conditions (Yang \textit{et. al.} \cite{Yang2021Research}), applications of machine learning-based cruise control systems (Farivar \textit{et. al.} \cite{Farivar2021SecurityOfNetwork}), V2X communications (Liu \textit{et. al.} \cite{Liu2020Economic},\cite{Failure2021ASafety}), and many more. Although, there are quite a few existing reviews on the ACC, they lack from several issues. These work do not address the challenges faced by the MPC algorithm and also do not provide a bird's eye view on the existing literature. In this regards, our review proposes a classification taxonomy, considering the general capabilities and building blocks of ACC, with focus on MPC. Also, we evaluate the most recent works (in the past five years) to identify the gaps and elucidate future directions necessary to address these gaps. Thus, our contributions in this manuscript are as follows.
    
\begin{itemize}
      	\item \textit{Background}- We provide a thorough explanation of ADAS and its associated functionalities \textit{i.e.,} ACC.
		\item \textit{Challenges}- We discuss the prominent challenges faced of ACC.
		\item \textit{Classification taxonomy}-We provide a detailed taxonomy to classify the existing works. With this, we portray the gaps in the current works.
		\item \textit{Thorough Review}- We have conducted an in-depth 	literature review of the existing works. We identify their pros and cons.
		\item \textit{Future Directions}- We describe the future directions that enables to optimize the functionalities of ACC.
\end{itemize}    	
    In this work, we conducted a thorough survey on the topic of ADAS and ACC. We collected papers from the journals of IEEE, ACM, Springer, Elsevier, \textit{etc.} We manually collected manuscripts by performing a keyword-based search using words ADAS, cruise control, ACC, \textit{etc.} In this process we collected around 100 papers of which 30 were rejected as they were out of our scope. Further, we classify them in our taxonomy\ref{sec:litsurvey}. In doing so, we outline various methodologies to identify the gaps in this field, and later propose future directions to fill these gaps.   

	The organization of rest of the paper is as follows. Section \ref{sec:back} offers a succinct overview of the introduction of ADAS and ACC. Section \ref{sec:litsurvey} is divided into two subsections, the first subsection describes our classification taxonomy. With the help of this, we perform our review of the existing literature. Section \ref{sec:dis} contains a comprehensive analysis of the reviewed literature, including a thorough examination of each category, discussing their merits, drawbacks, and associated challenges. Section \ref{sec:future} addresses the gaps identified in this domain and provides recommendations for the improvement of ACC. Finally, we provide our conclusion in Section \ref{sec:con}. 
	\par

%% file: includes/sections/background.tex
\section{Background}
\label{sec:back}

In this section, we present the essential principles of the ACC. This section is divided into four subsections. In the first sub section, we introduce ADAS. Then, in the second subsection, we delve into ACC which is the core of ADAS that brings innovative self-driving capabilites. Later in the third subsection, we describe the architecture of ACC by providing a diagrammatic explanation of its various components. Lastly, in the fourth subsection, we discuss the challenges faced by the ACC-equipped vehicles.
\par 

\subsection{Introduction to the ADAS}
Advance Driver Assistance System (ADAS) is designed to minimize or completely remove human errors in various types of vehicles. Vehicles equipped with ADAS feature an assortment of sophisticated sensors functioning as RADAR, SONAR, LiDAR sensors, \textit{etc.,} which enhance the driver's vision, hearing, and decision-making capabilities \cite{BACKCIT2}. Apart from the sensors, it is also embedded with interfaces and processors. ADAS is majorly classified in two ways as shown in the Figure \ref{fig:ADASTaxonomy}. Firstly, the passive ADAS systems, where the computer solely alerts the driver about a hazardous situation. It is the responsibility of the driver to take action to avoid the situation leading to an accident. Customary passive ADAS functions are an Anti-lock Braking System (ABS) used for skidding and turning when the emergency brake is applied, Electronic Stability Control (ESC) used to prevent both under or over-steering, Traction Control System (TCS), used to maintain a proper adherence during turns. The rear-view camera offers a perspective of what's behind the vehicle to the driver, while Lane Departure Warning (LDW) alerts when the vehicle is not keeping within its lane. Forward Collision Warning (FCW) informs the driver to brake so as to avoid a collision. Blind spot detection warns the driver if there is any vehicle in their blind spot and parking assistance guides the driver when their front or rare bumpers are approaching an object at low speeds \cite{ACCSIM}. The second type is the active ADAS system, under which lies our main scope of study \textit{i.e., ACC} which performs required navigation actions. A few of the facilities provided by ACC include automatic emergency braking used to avoid collision with other vehicles, emergency steering is employed to evade collision with an object in the lane, while lane keeping assist and lane centering guide the vehicle to remain centered within its lane. Additionally, traffic jam assist provides semi-automated help during stop-and-go situations, and self-parking is used to park the vehicle without the driver's assistance. In the subsequent sections, we focus on the topic of our study, \textit{i.e.,} ACC systems.
\input{./includes/tikz-figure/fig-2}

\subsection{Key Terms in Advanced Driver Assistance Systems (ADAS)}
The figure \ref{fig:Eg} illustrates key terms in Advanced Driver Assistance Systems (ADAS) using a visual representation of vehicles in different configurations. At the top, the lead car, shown in red, is the vehicle setting the pace and direction for the others. Below it, the ego car, depicted in blue, is the primary vehicle equipped with ADAS features and sensors that monitor the surrounding environment. To the right of the ego car is a string of cars (in green), representing a line of vehicles following each other in the same direction. Below the string, the homogeneous platoon (in orange) consists of identical vehicles traveling closely together, benefiting from synchronized movements and reduced aerodynamic drag. At the bottom, the heterogeneous platoon (in purple, yellow, and cyan) includes different types of vehicles, requiring advanced coordination due to their varying capabilities. 
\par 
For example, consider a highway scenario where a lead car is followed by an ego car equipped with adaptive cruise control. The ego car maintains a safe distance from the lead car while a string of following vehicles adjusts their speed based on the lead and ego cars. Further along the highway, homogeneous and heterogeneous platoons use vehicle-to-vehicle communication to travel efficiently as groups, improving traffic flow and fuel efficiency.
\input{./includes/tikz-figure/fig-5}

\subsection{Adaptive Cruise Control}
One of the crucial component of the ADAS system, as mentioned earlier, is Adaptive Cruise Control (ACC). It is engineered to assist vehicles in ensuring safe distances and speed limits to avoid any type of accidents \cite{Mutzenich2021Updating}. Sensor technology is used in which the car is embedded with cameras, lasers, and radar equipments. These help in creating an idea of how close the neighboring vehicles are to one another and the distance between the near objects to the vehicles is also calculated and displayed. There are six types of ACC. They are describes as follows.

\begin{enumerate}
	\item\textit{Radar-based System:} They operate by installing radar-based sensors either around or on plastic fascias to identify the vehicle's environment. The sensors emit radio waves or microwaves towards the road or vehicle ahead. The waves bounce back when an object is encountered. The radar sensor then detects the returning waves and calculates the distance concerning them \cite{Radar2024}. 
	\item\textit{ Laser-based System:} This system uses laser sensors, often referred to as Light Detection and Ranging (LiDAR). These sensors emit laser beams or pulses of light to detect objects. LiDAR provides high-resolution 3D mapping of the environment, allowing for precise detection. These laser beams bounce off objects in front of the vehicle, allowing the LiDAR sensor to detect and calculate the distance \cite{Lidar2024}.
	\item\textit{Three Binocular Computer Vision System:} This technology replicates human binocular vision by using two cameras to capture images simultaneously. These cameras are placed at a specific distance apart, mimicking the separation between human eyes. They capture images of the same scene from slightly different perspectives. The pictures are compared to determine the pixel differences between corresponding points in the left and right images. These disparities are then transformed into a 3D depth map, providing information about the distance \cite{Binvi2024}.
	\item \textit{Predictive System:}  Predictive Adaptive Cruise Control (PACC) proactively adjusts the vehicle's speed based on predictive data about traffic and road conditions. It relies on radar sensors to provide real-time data about the speed and distance, GPS, and map data to anticipate upcoming road features \cite{Pred2024}.
	\item \textit{Multi-Sensor System:} This technology utilizes a combination of various sensors to monitor the vehicle's surroundings \cite{Mulsen2024}.
\end{enumerate}
\input{includes/tables/sample-lit-comp-table8}

\subsection{Architecture of Adaptive Cruise Control}

The architecture of an ACC is as follows \ref{fig:Architecture}, firstly, sensors are placed on or in the vehicle to detect motion. They are of various types such as radar sensors that are primary sensors for ACC which uses radio waves to detect other vehicles. LiDAR sensors use laser beams to generate high-resolution data. Camera systems are used to capture images of the road and recognize lanes, signs, \textit{etc} \cite{ACCARCHITECTURE}. Ultrasonic sensors are used for low-speed maneuvering, and detecting obstacles. This information is passed on to the central processing component of the ACC system, also referred to as the Electronic Control Unit (ECU). It consists of various algorithms that are responsible for decision-making. It houses an algorithm for distance and speed control, following distance determination, speed adjustment, and object tracking. After the control unit has concluded its decision, the actuators convert this decision which is given in the form of an electric signal into action. The most commonly used actuators are the throttle actuator that adjusts the amount of fuel, the brake actuator that slows down the vehicle, and steering actuators used to assist with maintaining the vehicle within its lane \cite{BACKCIT4}. ACC also includes a Human Machine Interface (HMI) that allows the driver to interact with customized ACC settings. Thus, the architecture of an ACC combines sensors, control units, algorithms, actuators, and user interfaces. These components work together to improve both safety and driver convenience.

\input{./includes/tikz-figure/fig-1}

\subsection{Challenges of Adaptive Cruise Control}
Although ACC provides safety and comfort, it has many challenges, some of them are mentioned as below.
\begin{enumerate} 
	\item \textit{ Limited Awareness of Unpredicted Events ($C_{1}$)} : Caused due to the sensors that have difficulty predicting unpredictable events. 
	\item \textit{Poor Weather Conditions ($C_{2}$)} : Weather condition can affect sensor accuracy, leading to false readings. Examples of these conditions are heavy rains, sun, fog \textit{etc}.
	\item\textit{ Limited Effectiveness in Stop-and-Go Traffic ($C_{3}$):} It is due to the struggle taken by the ACC system to smoothly handle abrupt stops and acceleration. 
	\item\textit{ Undetectable Objects ($C_{4}$):} Inaccurate detection of stationary objects in hilly regions and curves.
	\item \textit{Sensor Obstruction ($C_{5}$):} Due to dirt, snow, or debris may hinder the performance of the ACC system.
	\item \textit{Driver Understanding and Misuse ($C_{6}$):} Occur by relying too heavily on the system.
	\item \textit{System compatibility and Integration ($C_{7}$):} It significantly contributes to the driver's safety. If not integrated properly, it can be a threat.
	\item \textit{Cost and Accessibility ($C_{8}$):} These factors may limit the use ACC system.
	\item\textit{ Legal and Regulation Challenges ($C_{9}$):} They must be considered including the question about liability in accidents.
	\item \textit{Cybersecurity ($C_{10}$):} It concerns that potentially compromise the safety and functionality of the system.
	However, ACC remains a valuable tool if the drivers understand the limitations of ACC.  
\end{enumerate}

\input{includes/tables/sample-lit-comp-table81}
Table \ref{tab:table81} outlines various types of ADAS categorized by their underlying technology and components. Radar-based systems rely on transmitter and receiver components to navigate through environmental conditions, facing challenges related to factors such as weather and interference. Laser-based systems, utilizing LiDAR sensors and optical receivers, must adhere to strict regulatory and safety standards, alongside coping with environmental variables. Binocular computer vision systems, driven by image sensors and processing units, face computational resource limitations while striving to interpret visual data accurately. Predictive systems, integrating radar and Lidar sensors, play a crucial role in decision-making and risk management for driving scenarios. Multi-sensor systems, combining GPS, camera systems, and radar sensors, are challenged by data processing time constraints as they synthesize information from various sources for comprehensive situational awareness. Each system type encounters specific challenges related to environmental factors, regulatory compliance, computational resources, decision-making, and data processing time, essential considerations in the development and implementation of advanced driver assistance technologies.

Table \ref{tab:table8} delineates various types of ACC and their associated components, factors affecting their performance and the challenges they encounter] Radar-based systems include components such as transmitters and receivers, influenced by environmental conditions and facing challenges C1, C3, C4, C5, and C8. Laser-based systems, comprising Lidar sensors, optical receivers, and signal processing units, are affected by regulatory and safety conditions, with challenges C1, C3, C4, C5, C7, C8, C9, and C10. Three binocular computer vision systems utilize image sensors and image processing units, requiring significant computational resources and encountering challenges C1, C3, C4, C5, C7, C8, C9, and C10. Predictive systems include radar and Lidar sensors, involved in decision making and risk management, facing challenges C1, C3, C4, C5, C7, C8, C9, and C10. Multi-sensor systems consist of GPS and camera systems, dealing with data processing time and facing challenges C1, C3, C4, C5, and C8.

%% file: includes/tikz-figure/fig-2.tex
\begin{figure*}[!h]
	\begin{forest}
		qtree,
		multiple directions={minimum height=4ex, anchor=center, forked edge}
		[ADAS%
		[, grow' subtree=east
		[Passive%
		[Anti-lock Breaking (ABS)]
		[Electronic Stability Control (ESC)]
		[Traction Control System (TCS)]
		[Lane Departure Warning (LDW)]
		[Forward Collision Warning (FCW)]
		[Blind Spot Detection]
		[Parking Assistance]
		]
		]
		[, grow' subtree=west
		[Active%
		[Automatic Emergency Braking]
		[Emergency Steering]
		[Lane Keeping Assist and Lane Centering]
		[Traffic Jam Assist]
		[Self Parking]
		[Adaptive Cruise Control (ACC)]
		]
		]
		]
	\end{forest}
	
	\caption{Different Components of the ADAS System}
	\label{fig:ADASTaxonomy}
	
\end{figure*}
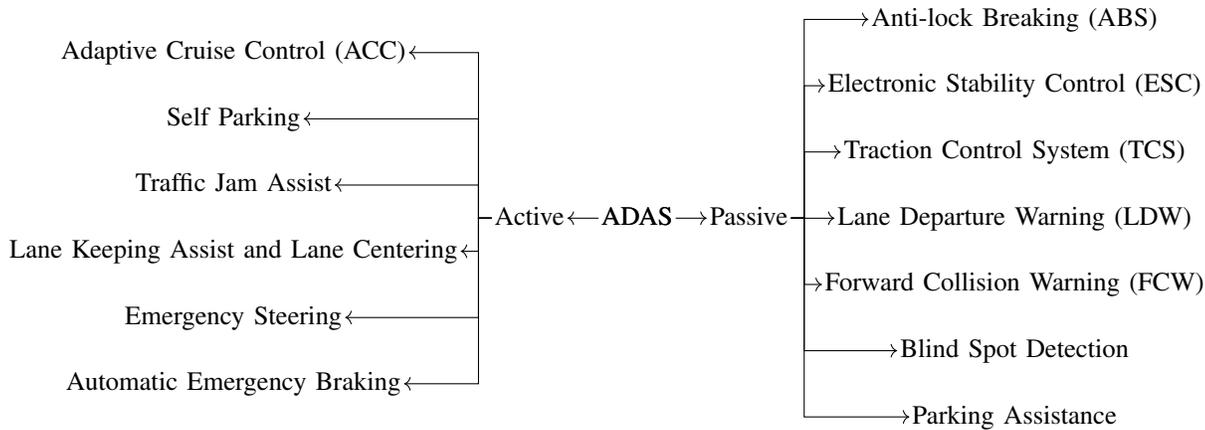

%% file: includes/tikz-figure/fig-5.tex
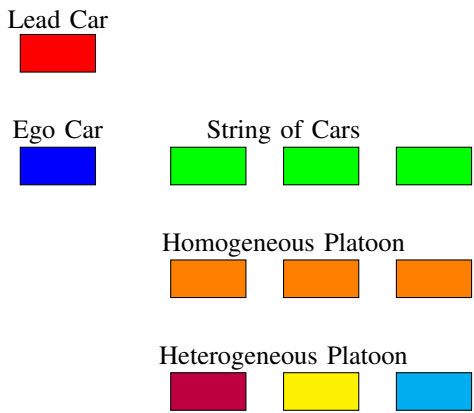
\begin{figure}
\begin{tikzpicture}
	
		\draw[fill=red] (1, 5) rectangle (2, 5.5);
		\node at (1.5, 5.7) {Lead Car};
		
		\draw[fill=blue] (1, 3.5) rectangle (2, 4);
		\node at (1.5, 4.2) {Ego Car};
		
		\foreach \i in {0, 1, 2} {
			\draw[fill=green] (3 + \i*1.5, 3.5) rectangle (4 + \i*1.5, 4);
		}
		\node at (4.5, 4.2) {String of Cars};
		
		\foreach \i in {0, 1, 2} {
			\draw[fill=orange] (3 + \i*1.5, 2) rectangle (4 + \i*1.5, 2.5);
		}
		\node at (4.5, 2.7) {Homogeneous Platoon};
		
		\foreach \i/\color in {0/purple, 1/yellow, 2/cyan} {
			\draw[fill=\color] (3 + \i*1.5, 0.5) rectangle (4 + \i*1.5, 1);
		}
		\node at (4.5, 1.2) {Heterogeneous Platoon};
\end{tikzpicture}
			\caption{Illustrations of key ADAS terms}
            \label{fig:Eg}
\end{figure}

%% file: includes/tables/sample-lit-comp-table8.tex
\begin{table*}
	\centering
	\caption{Different Types of the ACC}
	\label{tab:table8}
	\ra{1.8}
	\begin{tabular}{@{}
			m{0.2\textwidth}
			m{0.2\textwidth}
			m{0.2\textwidth}
			m{0.2\textwidth}
		}
		\midrule 
		\multicolumn{1}{c}{\textbf{Type}} &
		\multicolumn{1}{c}{\textbf{Components}} & 
		\multicolumn{1}{c}{\textbf{Implicators}} & 
		\multicolumn{1}{c}{\textbf{Challenges}} \\
		
		\toprule
		Radar-based systems & Transmitter, receiver \textit{etc.} & Environmental conditions & $C_{1}$,$C_{3}$,$C_{4}$,$C_{5}$,$C_{8}$\\  
		
		Laser-based systems & Lidar sensors, optical receiver, and signal processing unit \textit{etc.} & Regulatory and safety conditions &$C_{1}$,$C_{3}$,$C_{4}$,$C_{5}$,$C_{7}$,$C_{8}$,$C_{9}$,$C_{10}$\\
		
		Three binocular computer vision system &  Image sensors, image processing unit, \textit{etc.} & Computational resources & $C_{1}$,$C_{3}$,$C_{4}$,$C_{5}$,$C_{7}$,$C_{8}$,$C_{9}$,$C_{10}$\\
		
		Predictive Systems & Redar sensors and Lidar sensors \textit{etc.} & Decision making and risk management & $C_{1}$,$C_{3}$,$C_{4}$,$C_{5}$,$C_{7}$,$C_{8}$,$C_{9}$,$C_{10}$\\ 
		
		Multi-sensor Systems & GPS, camera systems \textit{etc.} & Data processing time & $C_{1}$,,$C_{3}$,$C_{4}$,$C_{5}$,$C_{8}$\\ 
		
		\bottomrule
	\end{tabular} 	
\end{table*}

%% file: includes/tikz-figure/fig-1.tex
\begin{figure}
	\centering
	
	\begin{tikzpicture}  
		
		\node[block] (n1) {\includegraphics[width=.3\textwidth]{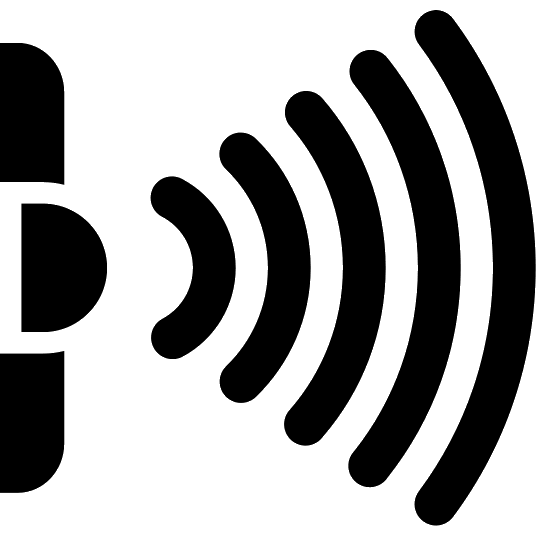}\\Sensors}; 
		\node[block,right=of n1] (b) {\includegraphics[width=.6\textwidth]{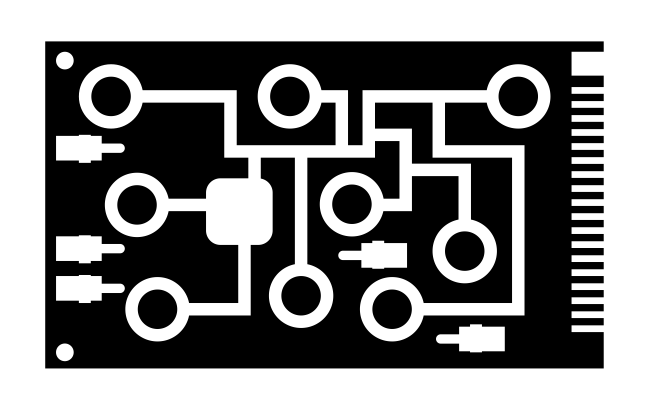}\\ECU};   
		\node[block,right=of b] (c) {\includegraphics[width=.6\textwidth]{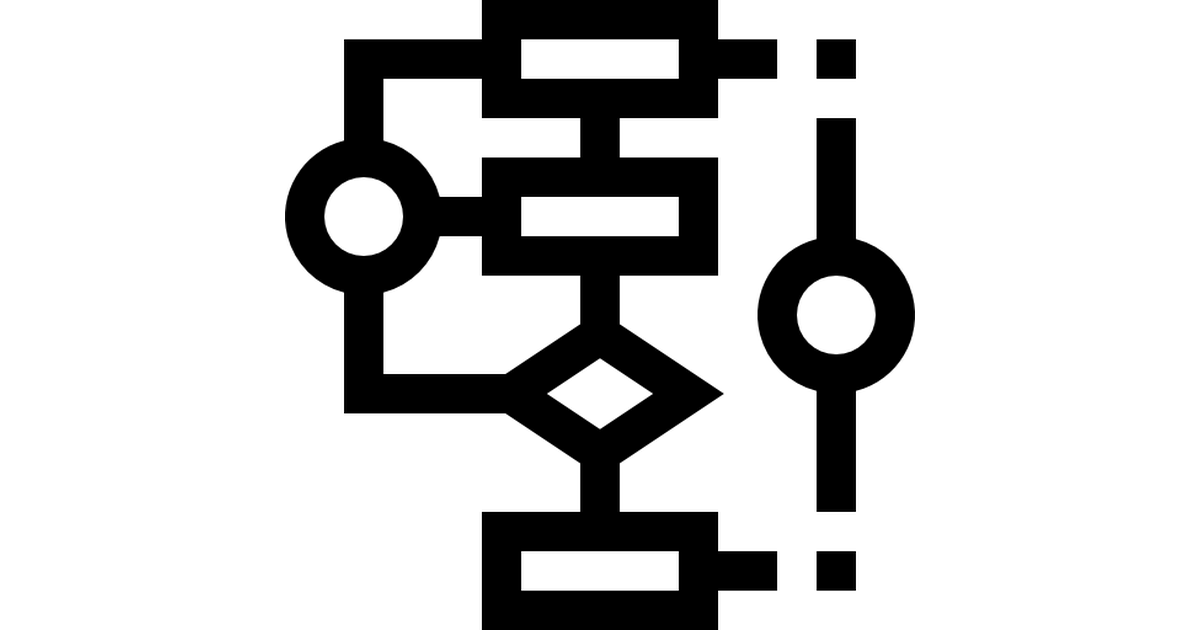}\\Algorithms};  
		\node[block] (d) at ([yshift=3cm]$(n1)!1.0!(b)$) {\includegraphics[width=.3\textwidth]{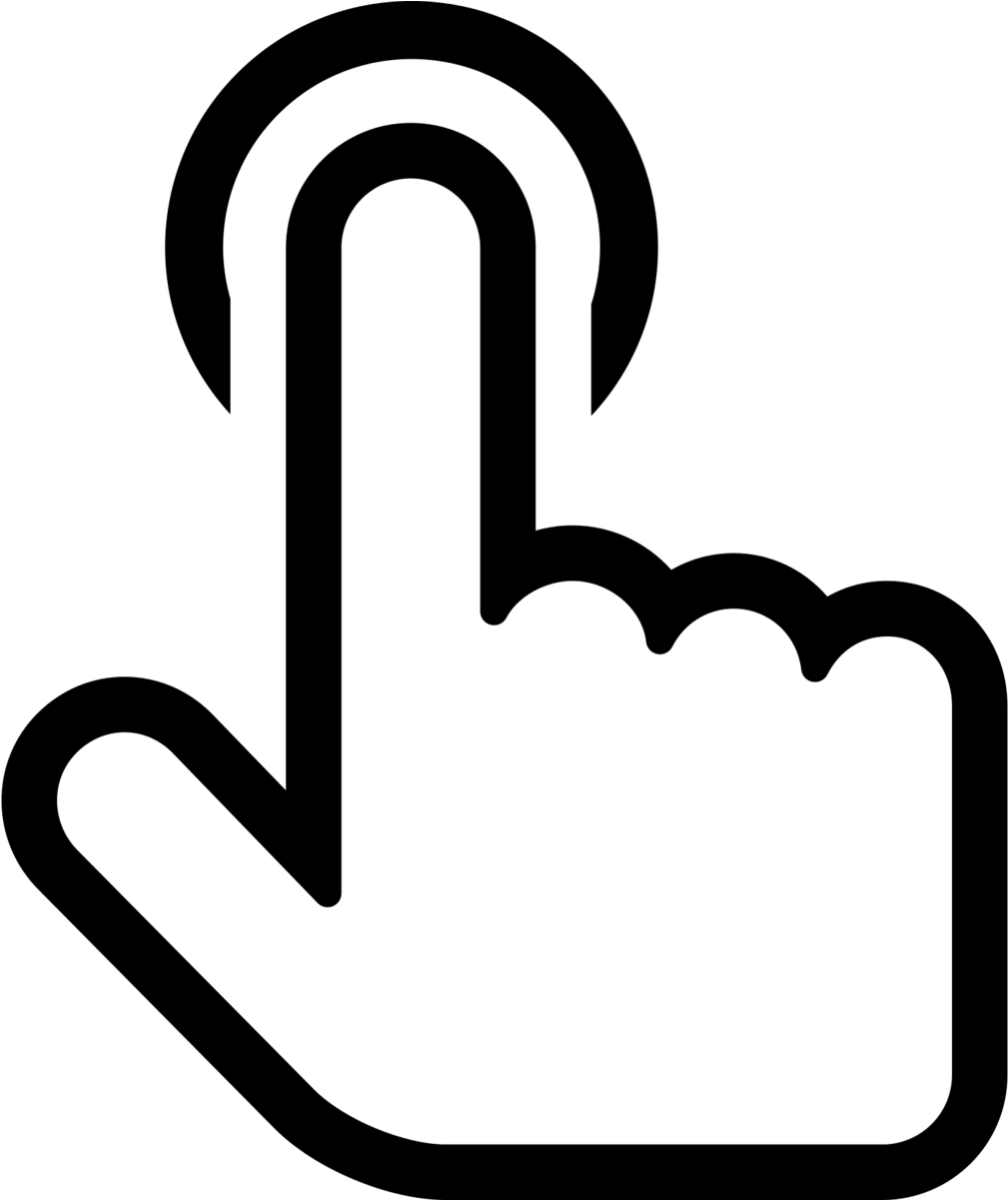}\\HMI};  
		\node[block] (e) at ([yshift=-3cm]$(b)!0.0!(c)$) {\includegraphics[width=.3\textwidth]{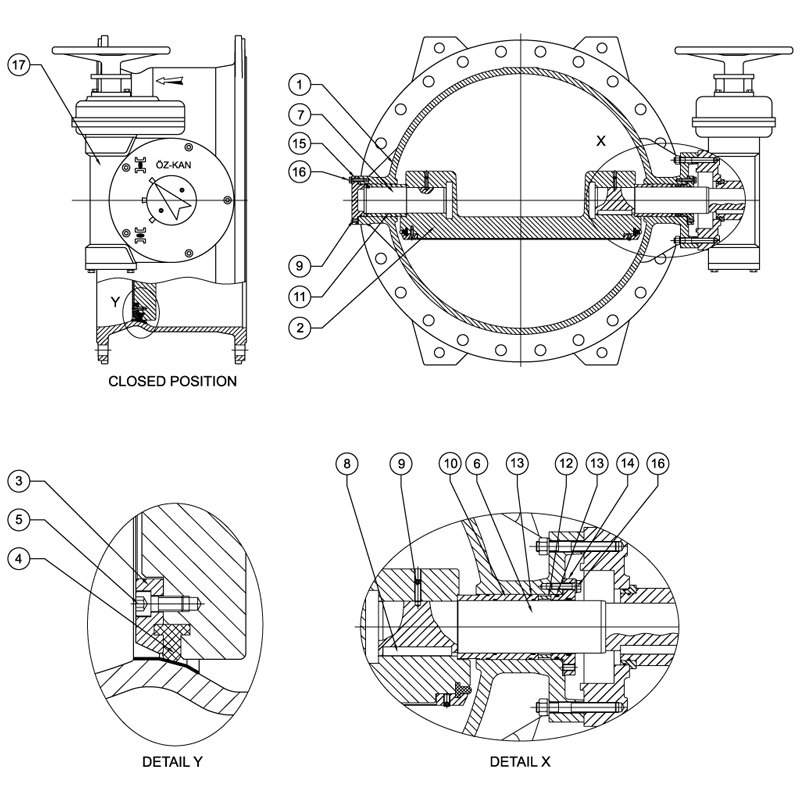}\\Actuators};  
		 
		
		\draw[line] (n1)-- (b);  
		\draw[line] (c)-- (b);    
		\draw[line] (b)-- (e);  
		\draw[line] (d)-- (b); 
				
	\end{tikzpicture}  
	\label{fig:Architecture}
	\caption{Architectural Components of the ACC}
\end{figure}
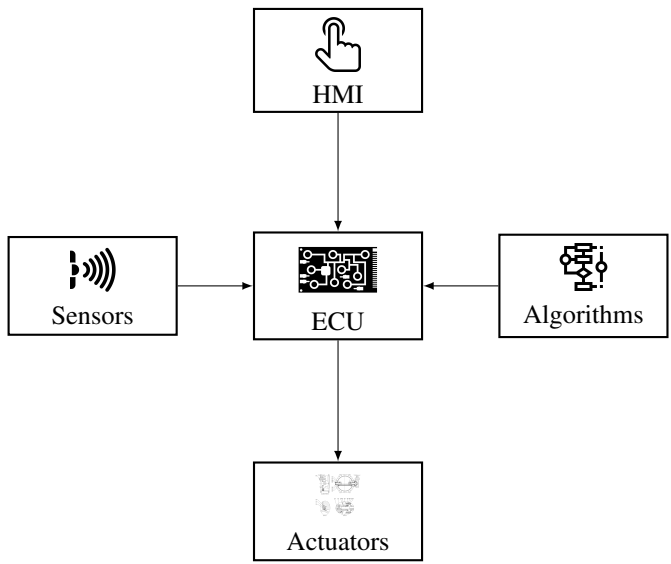

%% file: includes/tables/sample-lit-comp-table81.tex
\begin{table*}
	\centering
	\caption{Challenges and causes.}
	\label{tab:table81}
	\ra{1.8}
	\begin{tabular}{@{}
			m{0.05\textwidth}
			m{0.2\textwidth}
			m{0.2\textwidth}
			m{0.1\textwidth}
			m{0.15\textwidth}
			m{0.15\textwidth}
		}
		\midrule 
		\multicolumn{1}{c}{\textbf{Notation}} &
		\multicolumn{1}{c}{\textbf{Cause}} & 
		\multicolumn{1}{c}{\textbf{Description}} & 
		\multicolumn{1}{c}{\textbf{\makecell{Target \\ component}}} & 
		\multicolumn{1}{c}{\textbf{\makecell{Impact on\\ ACC}}} &
		\multicolumn{1}{c}{\textbf{Solution}} \\
		
		\toprule
		$C_{1}$ & Limited knowledge & Hampers ACC's ability to sense and respond & Sensors & Decreases accuracy of ACC & Sensor fusion technology \\
		
		$C_{2}$ & Poor weather conditions & Sensors & Reduced responsiveness & Weather resistant sensors\\
		
		$C_{3}$ & Limited effectiveness in traffic & Misinterpretation of data & Sensors & Increase complexity & Multi sensor fusion systems \\
		
		$C_{4}$ & Undetectable objects & Obstacles that system fails to detect & Sensors & Leads to collisions & advance sensor fusion technology\\
		
		$C_{5}$ & Sensor obstruction & Physical or environmental barriers & Sensors & Delayed or incorrect response & Advanced sensor cleaning mechanism \\
		
		$C_{6}$ & Driver understanding and misuse & Misunderstanding system capabilities & HMI & Improper use of ACC features & Driver education programs\\
		
		$C_{7}$ & System capabilities and integration & Capacity of interaction & Controller & Maximizing safety & Standardized communication protocols\\
		
		$C_{8}$ & Cost and accessibility & Adoption and implementation & Sensors, actuators and controllers & Prevalence and affordability & Cost effectve manufacturing method\\
		
		$C_{9}$ & Legal and regulation challenges & Navigating legal frameworks & Controller and HMI & Affects in real-world driving senarios & Standardized regulatory frameworks\\
		
		$C_{10}$ & Cybersecurity & Safeguard data & Controller & Compromise system integrity & Intrusion detection system\\
		 
		\bottomrule
\end{tabular} 	
\end{table*}

%% file: includes/sections/literatureSurvey.tex
\section{A REVIEW OF RESEARCH EFFORTS}
\label{sec:litsurvey}
In the previous sections we looked into the architecture of ACC and outlined its benefits and challenges. In this section, we review the existing literature in the field of cruise control. For this we divided the section into two sub sections. The first subsection provides a classification taxonomy of the existing works. The second subsection provides a overview of the state-of-the-art works present in this field. 
\subsection{Overview}
\label{sec:sclass}
In this subsection, we present our classification taxonomy of the reviewed literature and categorize them into seven broader categories \textit{i.e.,}. 
\begin{itemize}
	\item \textbf{Safety}: This category includes studies focused on ensuring the safety of systems and users. It covers aspects such as accident prevention, safety protocols, and risk assessment.
	\item \textbf{Techniques}: This encompasses various methodologies and approaches used in the field. It includes algorithm development, system design, and implementation strategies.
	\item \textbf{Security}: This category addresses the protection of systems against threats and vulnerabilities. It includes encryption methods, intrusion detection systems, and secure communication protocols.
	\item \textbf{V2X (Vehicle-to-Everything)}: This involves communication between vehicles and other entities such as infrastructure, pedestrians, and networks. It includes studies on V2V (Vehicle-to-Vehicle), V2I (Vehicle-to-Infrastructure), and V2P (Vehicle-to-Pedestrian) communication.
	\item \textbf{Energy Conservation}: This category focuses on strategies to reduce energy consumption and improve efficiency. It includes power management techniques, energy harvesting, and sustainable practices.
	\item \textbf{Human Factors}: This involves the study of how humans interact with systems. It includes user interface design, ergonomics, and the impact of human behavior on system performance.
	\item \textbf{ML-based (Machine Learning-based)}: This category includes studies that apply machine learning techniques to solve problems in the field. It covers areas such as predictive modeling, data analysis, and autonomous decision-making.
\end{itemize}
We sub-categorize our taxonomy by being more specific as depicted in Figure \ref{fig:Taxonomy}. This helps us to gain an in-depth analysis of the topic and understand the gaps in this domain.

\input{./includes/tikz-figure/fig-3}

\subsection{A Review of Existing Works}
\label{sec:srev}
In this subsection we perform a thorough analysis of the existing works in this field, we do so based on the classification taxonomy presented in the previous subsection. We pertain various techniques implemented by the previous works in achieving a stable ACC. We also give their advantages and shortcoming.
\label{sec:srev}

\subsubsection{Safety}
ACC directly translates the concept of general safety into driving. Traditional cruise control maintains a set speed, but ACC adds a crucial layer of protection. It uses sensors to constantly regulate the surrounding environment. If the lead vehicle slows down, ACC reacts by automatically adjusting the ego car's speed to maintain a safe distance. This not only minimizes the risk of accidents, but by reducing the need for constant vigilance and reaction time, it also helps to alleviate driver stress and fatigue.
Chen \textit{et. al.}, \cite{Chen2021Economic} proposed a strategy for increasing the life span of batteries in EVs using the eco-ACC strategy. The proposed technique is independent of models, operates in real-time, and remains resilient in car-following situations. However, the proposed strategy lacks in control performance for connected and automated vehicles.
\par 
Wang \textit{et. al.}, \cite{Wang2022AFramework} presented a framework of safety analysis. Altrarica 3.0 is an advanced modeling language utilized for analyzing the safety of the automated vehicle. They have analyzed an approach that transforms the failure behavioral model into a Temporal Fault Tree (TFT). However, TFT needs significant manual labor involved, leading to high costs and time expenditure. Regardless of how, after compiling the algorithm it is seen that the obtained data is consolidated and integrated into a TFT system. It has multiple state results that makes it more efficient.
Gunter \textit{et. al.}, \cite{Gunter2020Model-Based} described a model-based string stability of the ACC system using field data. Experimentation is conducted using a luxury electric sedan equipped with a commercially available ACC system. String stability is assessed and it is found that the best-fit model is unstable. However, their primary contributions are creating a data-oriented method for determining string stability. But, the advancements for improvising the commercials ACC were not discussed.
\par 
Mu \textit{et. al.}, \cite{Mu2021StringStability} proposed a desired safety margins model for rear-end collision avoidance. Simulation findings showcase that the suggested algorithm enhances the smoothness of traffic flow and prevents collisions by determining an appropriate velocity for car-following scenarios. However, the proposed algorithm has only been investigated and has not been applied to real-life scenarios. 
Wang \textit{et. al.}, \cite{Wang2018Cooperative} proposes a strategy for optimizing the Flow Topologies (IFT) for ACC, labeled as the CACC-OIFT strategy. This approach is employed to enhance string stability in typical traffic conditions by dynamically toggling the "send" functionality. CACC-OIFT comprises an IFT optimization model alongside a Proportional-Derivative (PD) controller. Evaluation using NGSIM field data demonstrates that this strategy effectively improves string stability in V2V communication. However, scenarios of communication failures are not considered in this system.
\par 
Lu \textit{et. al.}, \cite{Lu2018NewAdaptive} proposed a Smart Driver Model (SDM) to describe the autonomous vehicle flow. It’s an ACC that aims stability of homogeneous traffic flow. After implementation and testing, it is seen that this model is consistent with the result of linear stability analysis and stabilizes traffic flow. Nevertheless, the analysis is theoretical, lane-changing behavior is ignored and different traffic conditions with ACC embedded vehicles are not considered.
Arnaout \textit{et. al.}, \cite{Arnaout2014Progressive} came up with a progressive deployment strategy for ACC to improve traffic dynamics. They have used previously developed traffic simulation models. The motorway model is designed to induce discomfort and trigger stop-and-go traffic. The experiment mentioned above demonstrates a decrease in traffic congestion. Allowing ACC vehicles access to high occupancy vehicle lanes could potentially enhance highway capacity with a ACC penetration rate of 40\%. However, additional cases could be added while experimenting, like allowing accidents to occur and facing vehicles to be traveled on smaller lanes \textit{etc.,} to explore the impacts of ACC in dynamics traffic.
\par 
Vahidi \textit{et. al.}, \cite{Vahidi2003ResearchAdvances} conducted a study on collision avoidance systems. Their main focus is on particular areas like avoiding collisions, influence on the comfort of drivers, safety, and the flow of traffic. It is discussed that advanced highway systems have a lot of financial, technical, and institutional barriers, unlike vehicle-based assist systems. They also explain how an ACC system improves driver’s comfort, which researchers had two perspectives. First, ACC helps in reducing the driver workload, and second, the poor design of ACC can be hazardous. However, they did not focus on determining appropriate following distance for different drivers that is a crucial part in avoiding collision.
Elmorshedy \textit{et. al.}, \cite{Elmorshedy2022Quantitative} conducted a study on the impacts of time headway in car following model. The Intelligent Driver Model (IDM) and Shladover’s model are compared using Aimsun micro simulation. Simulation results show that IDM has a low response. Adding on to the analysis, an on-off ACC-based strategy for optimizing the traffic is depicted. It aims to enhance the freeway performance. After testing, it is seen that this strategy improves average throughput and speed. However, the focus is on the impacts of urban traffic rather than proposing a better ACC system and it is relied on the assumption that the ACC systems remain unchanged in the future.
\par 
Calvert \textit{et. al.}, \cite{Calvert2020Cooperative} contributes a field operational test on the real traffic. It demonstrates ACC vehicle's ability to perform with lower time-headway conditions. Platoon dissolvement and cut-ins are analyzed and it is seen that the ACC operation are demonstrated with frequent re-coupling of platoons. However, due to its limited penetration rates optimized traffic flows improvement cannot be derived.
Milanes \textit{et. al.}, \cite{Milanes2014Cooperative} designed a improvised ACC system. It has two controllers, first one is used to lead the vehicle and the other to supervise car-following after the vehicle joins the convoy. They have implemented it in M56 vehicles equipped with DSRC devices and are tested on public roads. The test cases included reducing gap variables and handling non-equipped vehicles. The response time and string stability of ACC are better compared to other. The work does not focus on traffic responses of the ACC.
\par 
Shladover \textit{et. al.}, \cite{Shladover2012Impacts}, contributed to improving the market penetrations of ACC on highway capacity. They made use of the Aimsum Microscopic simulator site with four different types namely, manual vehicle, ACC, HIA, and CACC. Although, the result indicated that ACC is not likely to generate any meaningful difference in the capacity of highways as users are using ACC at gap setting equal to a huge positive impact only in the cases, where all vehicles are Co-operaative Adaptive Cruise Control (CACC) embedded, the maximum lane capacity would be about 4000 vehicles per hour.
Melson \textit{et. al.}, \cite{CLMelson2018Dynamic} analyzed the impacts of CACC on density of trafic flow and designed a model for dynamic traffic assignment. To study the effects of CACC on larger networks, they incorporated CACC into the model. First, with the help of the MIXIC car-following model, the flow-density relationship is calculated. It is then followed by applying the relationship in system. The results indicate that they reduce freeway congestion, but resulted in the increase of the travel time.
\par 
Hu \textit{et. al.}, \cite{Hu2020Cooperative} proposed an adaptive leader-following approach. The strategy employed involves a dual-layer distributed control system designed to uphold the string stability of diverse and interconnected vehicle convoys traveling in unison with a fixed spacing protocol. The outcomes demonstrate the effective functioning of the convoy control method even when faced with disruptions from the lead vehicle. However, challenging situations, such as uneven roads, hardware malfunctioning, \textit{etc.,} are not discussed.
Gunter \textit{et. al.}, \cite{Gunter2021Commercially} enhanced the string stability of ACC systems that are in use commercially. They conducted stability tests on seven vehicle models from two manufacturers. By utilizing car-following techniques and analyzing data gathered from over 1,200 miles of driving, they developed delay differential equation models. It was observed that the string of all vehicles tested was unstable. Nevertheless, there remains the potential for commercial ACC systems to surpass human drivers.
\par 
Pangwei \textit{et. al.}, \cite{Wang2014AnImproved} contributed in the upgradation of the CACC algorithm grounded in sliding mode control theory. The suggested algorithm proceeds through three key steps. Initially, it establishes a replica of the deviation in vehicle spacing within a platoon. Secondly, string stability cases are reviewed and lastly, five vehicle models are used to enhance the CACC algorithm in MATLAB/Simulink. The CACC controller adjusts itself to maintain string stability and avoids chain collision accidents caused by partially invalid communication. It is feasible and covers the shortages of traditional CACC algorithms. The major drawback is that the controller must be robust and accurate.
Moon \textit{et. al.}, \cite{Moon2009Design} presented an algorithm that contributes to collision avoidance of full-range ACC system. This operates across three distinct modes: comfort, large-deceleration, and severe-braking. Test results indicate that the proposed algorithm delivers smooth tracking performance, both at high and low speeds. Additionally, it ensures a safe distance from the lead vehicle. However, various factors such as road conditions, strategy tuning, value tuning, \textit{etc,} are not studied in this work.
\par 
Wasserburger \textit{et. al.}, \cite{Wasserburger2020Probability} came up with a probability-based short-term velocity prediction method. This method relies solely on historical velocity measurements. The approach is integrated into a model predictive control algorithm designed for an ACC system tailored for heavy-duty vehicles. It is seen that the energy savings are up to 17\%. Its main advantage is that this method saves more energy compared to Markov chain-based prediction. However, the proposed work does not consider communication loss as a factor that hinders performance.
Wang \textit{et. al.}, \cite{Wang2021Online} contributed to ACC with online parameter estimation methods. Two online methodologies are employed for real-time system identification. The first employs the recursive least squares method, while the second tackles a nonlinear joint state and parameter estimation challenge using particle filtering. These approaches achieve the lowest mean absolute error in velocity and space gap, registering at 0.24 m/s and 2.02 m, respectively. This translates to errors of 0.8\% in velocity and 5.0\% in space gap. While scalable and suitable for real-time applications, efforts to further minimize space gap errors are warranted.
\par 
Mintsis \textit{et. al.}, \cite{Mintsis2021Enhanced} improvised in speed advice for connected vehicles while ensuring energy and traffic efficiency. Thessaloniki’s (city in Greece) microscopic model is used as a proving ground under various traffic conditions. The test has resulted in the speed advice being safe and comfortable to use. But the advice given depends on many factors like characteristics of the roadway, traffic signals, \textit{etc.} Hence, the implementation strategy of eco-driving significantly influences both traffic efficiency and environmental advantages.
Diba \textit{et. al.}, \cite{Diba2014Optimized} deliberated on optimizing robust cruise control systems for EVs. A resilient cruise control system is essential for ACC as it ensures accurate speed tracking and adherence to a predetermined vehicle speed. This system comprises two layers, the initial layer involves a speed control mechanism constructed around a proportional-integral-derivative controller, while the subsequent layer employs a torque control system implemented through a proportional-integral controller. After the simulations, it is seen that the cruise controller is robust and has disturbance rejection behavior that is satisfactory in performance. 
\par 
Magdici \textit{et. al.}, \cite{Magdici2017Adaptive} proposed an architecture that addresses the problem of following a vehicle with varying acceleration in a secured manner. This system comprises both a safety controller and a nominal controller. It has been evaluated using real traffic data, demonstrating excellent performance in tracking position and velocity while ensuring safety and comfort. However, future work could be done to improve the performance by upgrading the architecture.
Yang \textit{et. al.}, \cite{Yang2021Research} proposed an ACC system architecture with adaptability to multiple operating conditions. The optimization focuses on enhancing the functional requirements of the ACC system across varied road conditions. A cost-effective ramp cruise control strategy is devised for vehicles, leveraging dynamic programming theory. Additionally, a curve radius prediction algorithm is employed to ensure lateral safety, with validation conducted using a virtual simulation platform. However, all conditions were not considered for real-life testing.
\par 
Chen \textit{et. al.}, \cite{Chen2020Data} proposed a data-based parameter setting method. Extensive road tests were conducted, leading to the development of a method to evaluate the safety performance of ACC systems. Parameters such as jerk limit and time delay were thoroughly examined. This method offers detailed guidance and a systematic approach to assessing the safety performance of ACC systems. However, other complex parameters such as target braking are not considered.
Makridis \textit{et. al.}, \cite{Makridis2020Response} came up with a strategy to assess the response time of the controller in an ACC-equipped vehicle during car-following scenarios. Testing results display the ACC response time between 0.8 and 1.2 seconds which is close to human reaction time. Hence, ACC can be used especially in stop-and-go scenarios. However, various vehicles were not considered during testing which resulted in various unexplored automated functionalities.
\par 
Manolis \textit{et. al.}, \cite{MANOLIS2020102617} contributed to improving the motorway traffic. ACC-based traffic control strategy uses dynamic adaption of driver behavior of ACC-equipped vehicles whenever required. The findings indicate that as the penetration rate of ACC vehicles increases, there is a corresponding enhancement in traffic conditions. This strategy reduces fuel consumption. But, the results obtained were not accurate even after bounds were applied to certain cases.
In conclusion, the implementation of ACC heralds a significant stride forward in automotive safety across various dimensions. By intelligently adjusting vehicle speed to maintain safe distances from other vehicles, ACC substantially reduces the risk of collisions, particularly in scenarios characterized by traffic congestion or sudden changes in speed. Moreover, its ability to enhance driver comfort by reducing the need for constant speed adjustments can contribute to mitigating driver fatigue, further bolstering safety. However, while ACC holds immense promise, its efficacy hinges on comprehensive considerations of safety beyond collision avoidance. This encompasses aspects such as system reliability, cybersecurity safeguards, and user education to ensure optimal utilization and minimal risks of malfunctions or misuse. Additionally, ongoing advancements in sensor technology and vehicle-to-vehicle communication systems promise to fortify ACC's capabilities, thereby ensuring safety in diverse driving environments. As ACC continues to evolve, a holistic approach encompassing technological innovation, regulatory standards, and driver awareness is imperative to unlock its full potential in enhancing overall road safety.
\par 
\subsubsection{Techniques}
In the domain of automotive innovation, ACC emerges as a pinnacle of technological advancement, integrating an array of techniques across multiple dimensions to redefine driving experiences. At its core, ACC operates on a sophisticated fusion of sensor technologies, including radar, liDAR, and cameras, enabling real-time monitoring of surrounding traffic dynamics. These sensors meticulously gauge the distances and relative velocities of vehicles, facilitating seamless adjustments in speed to maintain safe inter-vehicle gaps. Moreover, ACC harnesses the power of machine learning algorithms to predict and adapt to diverse driving scenarios, enhancing its responsiveness and adaptability on the road. Beyond sensor fusion and algorithmic sophistication, ACC relies on robust communication protocols, facilitating seamless interaction between vehicles for coordinated traffic management. As we embark on an exploration of ACC techniques, it becomes evident that their convergence not only empowers drivers with enhanced safety and convenience but also heralds a paradigm shift towards intelligent and interconnected transportation ecosystems.
Zhu \textit{et. al.}, \cite{Zhu2020Synthesis} proposed a feed-forward strategy. Using this strategy, the synthesis of CACC is conducted. The control synthesis is of two types, one of which uses ACC feed-forward and the other uses control feed-forward. The results show are improvment in tracking performance and reducing design effort. However, it is limited to only homogeneous platoons.
\par 
Jia \textit{et. al.}, \cite{Pan2022Energy-Optimal}  designed a linear and non-linear Model Predictive Control (NMPC) that is implemented for energy-optimal Adaptive Cruise Control (EACC) in electric vehicles, utilizing a combination of time-domain Linear MPC (LMPC) and spatial-domain NMPC formulations. Comparative analysis reveals the significant advantages of NMPC over LMPC. 
Wu \textit{et. al.}, \cite{Wu2019Cooperative} proposed an algorithm with the Kalman filter that computes the acceleration of the leading vehicle, which is then relayed to the ego-vehicle's CACC system in the event of communication disruption. Mobile robots are employed to emulate driving scenarios. Findings indicate that the adaptive Kalman filter significantly outperforms existing methodologies when communication is lost. However, real-time implementation of the proposed technique has not yet been conducted and issues like platoon stability are not discussed.
\par 
Miyata \textit{et. al.}, \cite{Miyata2010Improvement} contributed to the improvement of ACC performance by creating a system with the following control that considers the vehicle’s slip side. For this, they have focused on two major fields. Initially, the configuration of the ACC system encompasses controls for maintaining constant velocity, deceleration, following distance, and acceleration. Secondly, ACC-ECU collects preceding vehicle information that is transmitted by millimeter wave radar. After testing, the preceding vehicle's lock-on performance on expressing sharp turns in mountainous regions, it turns out that the system reduces the burden on the driver by sufficient lock-on, speed up, speed down performance, and satisfactory driving experience for the driver. Yet, performance optimization schemes are not mentioned.
Milanes \textit{et. al.}, \cite{VMilanes2014Modeling} contributed to the development of the ACC and CACC control system. Four vehicles equipped with three different controllers are used for the experiment. The three controllers are production ACC, intelligent driver model, and newly developed CACC controller. After comparing it is found that the newly developed ACC and CACC systems match the experimental results very closely. Nevertheless, the impact of the model on traffic flow and the reactions of unequipped vehicles during cut-ins and cut-outs are not taken into account during testing.
\par 
Hidayatullah \textit{et. al.}, \cite{Hidayatullah2021Adaptive} proposed an ACC controller employing the gain scheduling method, that is employed to address model vehicle dynamics represented as a linear parameter varying system. For simulation, PreScan is used, as it provides high non-linear vehicle cases, and MATLAB/Simulink is used for decision-making and target tracking. The vehicle's mass is varied depending upon the number of passengers getting onto or down the vehicle. By using disk margin the algorithm assures robustness for various frozen points and rate changes, but discussions about string stability under differing conditions in vehicles with CACC are not addressed. 
Lunze \cite{Lunze2020Design} proposed a design of the communication structure of the CACC. The objective is to identify local vehicle controllers and communication frameworks while examining the circumstances under which control vehicles adhere to an asymptotic time-headway spacing policy. Test scenarios are designed to minimize individual vehicle delay and ensure platoon positivity externally. The findings demonstrate the feasibility of acquiring data communication from vehicles to their subsequent counterparts. However, a few of the factors that affect the design of feedback controller were not considered during the experimentation.
\par 
Feng \textit{et. al.}, \cite{Feng2021Robust} developed a resilient platoon control system suitable for mixed traffic flow, employing tube model predictive control as its foundation. Through numerical experimentation, the effectiveness of the approach is confirmed, showcasing its ability to manage uncertainty with reduced communication and computational overhead. However, the integration of prediction models can be further improved.
Brugnolli \textit{et. al.}, \cite{Brugnolli2019Predictive} designed the inner loop of ACC utilizing two separate model predictive control techniques, specifically finite horizon and infinite horizon prediction. The designed controllers communicate directly with the customized ECU. After simulations, it was seen that the controller satisfactorily tracked the changing speeds and maintained a safe gap from the leading vehicle. However, the enhancement of the controller performance is not discussed.
\par 
Ma \textit{et. al.}, \cite{Ma2020Cooperative} proposed a CACC strategy for EV platoons to enhance safety. They have used MPC based on the swarm optimization (SA-PSO) algorithm. It effectively addresses the optimization problem of nonlinear multi-objectives. Four EVs are used to compare results with the ACC strategy. Upon comparison, it becomes evident that the CACC strategy consistently maintains superior distancing, minimizes spacing, and exhibits a rapid response despite time delays. Additionally, it boosts the regenerative braking energy of the platoon by approximately 16.5\%, thus delivering economic advantages to drivers. However, there is little vibration in acceleration. Hence, to balance comfort and economy the SA-PSO algorithm must be optimized.
Zhang \textit{et. al.}, \cite{Zhang2021Data-Driven} developed an optimal forward-looking distributed CPS application tailored for safety-following driving control, facilitating the seamless development and adjustment of control models using cloud-based historical data for enhanced vehicle safety. After simulation, it is seen that this framework is feasible, effective, and improves cruise safety. However, the results also showed an unknown delay that affects the accuracy of the vehicle control, and because of the restricted test equipment availability, they refrained from developing an actual cloud system and a 5G high-speed communication system.
\par 
He \textit{et. al.}, \cite{He2021Defensive} developed an algorithm to minimize the time spent in Blind Spot Zones (BSZ) for self-propelled vehicles by taking neighboring vehicles into account. While many contemporary vehicles feature blind spot detection systems, a significant portion lack ADAS. Moreover, these systems fail to indicate the position of the vehicle they're installed in. Following demonstration across 100 scenarios, it was observed that the average dwelling time in neighboring vehicles' BSZs reduced by 46.3\%, resulting in decreased fuel consumption. These findings were derived from practical car-following simulations employing real-world traffic data. As, this algorithm uses an MPC controller to extract data about nearby vehicles’ information to predict the future speed, a small error in the prediction of speed can adversely affect the algorithm's performance.
Therefore, ACC embodies the core principle of safety by proactively identifying and mitigating hazards to create a safer driving experience.
\par 
\subsubsection{Security}
The widespread adoption of ACC in modern vehicles has revolutionized driver assistance capabilities. However, with this technological advancement comes a new set of security concerns.  ACC relies on communication and sensor data to function, introducing potential vulnerabilities that could be exploited by malicious actors. This raises critical questions regarding the security architecture of the ACC systems and the potential consequences of a successful cyberattack.
Lin \textit{et. al.}, \cite{Lin2020Robust} proposed a Robust Model Predictive Control (RMPC) strategy for linear time-invariant systems. The key feature of this approach lies in the segregation of information concerning disturbances in optimization, addressing robustness and performance individually. The selection of a quadratic Lyapunov function as the objective function may impose constraints on the flexibility and scope of RMPC. Nonetheless, this limitation is mitigated by introducing an explicit Lyapunov function that guarantees stability.
\par 
Lee \textit{et. al.}, \cite{Lee2021Design} developed CACC system incorporating an unconnected vehicle scenario. When encountering an unconnected vehicle, the CACC with Unconnected vehicle (CACCU) system in the loop establishes communication with a connected vehicle positioned further ahead. To enhance string stability, a speed-command-based CACCU controller is devised. This experiment employs two automated vehicles equipped with mobility sensors and Wi-Fi connectivity. Through six conducted tests, it was observed that CACCU achieved a reduction of 10.8\% in acceleration, a 60\% decrease in spacing error, and a 6.2\% decrease in fuel consumption, effectively averting traffic disturbances. However, generalized traffic scenarios are not considered in this work.
Sawant \textit{et. al.}, \cite{Sawant2021Robust} designed a disturbance observer based on sliding mode control for CACC in a platoon. By simulating various traffic scenarios, the scheme is verified. The findings indicate that the scheme effectively handles uncertainties in actuator dynamics, demonstrating robustness. It performs satisfactorily for both homogeneous and heterogeneous platoons, utilizing on-board sensors to gather immediate predecessor vehicle information. In the future, the scheme could use multiple predecessor information to increase its robustness.
\par 
Yu \textit{et. al.}, \cite{Yu2022Safety} explored the safety implications of CACC vehicle degradation amidst persistent communication disruptions. The study involved categorizing vehicles into four types: CACC, ACC, manually operated, and enhanced manually operated vehicles equipped with information-transmitting devices. Simulation findings indicate that manually operated vehicles demonstrate a capacity to mitigate rear-end collision risks during communication breakdowns. However, the integration of information-transmitting devices into manually operated vehicles appears to compromise their resilience to communication interruptions.
Zhang \textit{et. al.}, \cite{Zhang2020ControlDesign} contributed to a delay-compensating CACC controller. They focus on three areas that highlight the merits of delay-indemnifying CACC. Firstly, it ensures local stability and string stability by minimizing communication delays and time intervals. Secondly, in comparison to standard CACC, delay-tolerant CACC enhances local stability, string stability, and traffic flow stability. Lastly, owing to its reduced time intervals, delay-tolerant CACC enhances throughput and mitigates the impact of traffic disruptions. Apart from communication delays, sensor and actuator delays are two other types of delays to consider. Their effects and recoup are not mentioned at all.
\par 
Harfouch \textit{et. al.}, \cite{Harfouch2018AdaptiveSwitched} proposed an adaptive switched control method for managing heterogeneous platoons in the presence of inter-vehicle communication failures. This approach integrates a baseline controller with an adaptive term that operates in a switched manner. Upon analysis, it's observed that the controller ensures continuous communication, asymptotically driving the error towards zero, and upholds string stability even in the absence of communication. This approach was divided into two theories in which the result of theorem one holds under the assumption of ideal continuous communication.
Alotibi \textit{et. al.}, \cite{Alotibi2021Anomaly} proposed a kinematic model for anomaly detection for CACC. Their attention was directed towards a critical risk scenario involving the compromise of the platoon leader, which could result in traffic instability and collisions. The model they introduced enables vehicles and fixed infrastructure to detect and exchange information regarding platoon leaders, thereby enhancing reliability. Furthermore, this method demonstrated the capability to detect over 92\% of falsified data with fewer than 13\% false alarms during testing. However, the impacts of communication delay and different attack scenarios were not considered.
\par 
Xing \textit{et. al.}, \cite{Xing2020Compensation} contributed to mitigating communication delays within a homogeneous CACC system by employing a master-slave control strategy. This approach involves reordering communication delays to facilitate the utilization of a Smith predictor. Experimental testing reveals that the strategy ensures both individual vehicle stability and maintains string-stable platoons with time gaps of less than 0.10 seconds, regardless of the presence of communication delays. However, future improvements can be made in decreasing the string-stable time gap. 
Zhang \textit{et. al.}, \cite{Zhang2020ActiveFaultTolerant} developed an active fault-tolerant control approach for ACC systems, addressing occasional malfunctions in wave radar sensors that can compromise system safety. Utilizing a mixed logical dynamical model, the top segment of the control system integrates fault-free and fault dynamics. Within the model predictive control framework, an active fault-tolerant control model is implemented, ensuring secure and seamless driving unaffected by radar sensor failures.Thus, improving the vehicle’s intelligence, However, efforts are not made to make a tolerant-free system. 
\par 
Cui \textit{et. al.}, \cite{Cui2022Development} improved the robustness of CACC with dynamic topology, targeting the mitigation of delayed responses in unforeseen circumstances. Employing both all-predecessor-following and predecessor-leader-following control methods, the development controller strives to enhance responsiveness. The outcomes validate the proposal controller's string stability and robustness, affirming its effectiveness. The complexity and the computation time of the controller compromise its advantages.
Zhang \textit{et. al.}, \cite{Zhang2023Adaptive} proposed an adaptive Radial Basis Function (RBF). This system relies on an auxiliary sensor to manage nonlinear platoons through vehicular ad-hoc networks amidst Denial-of-Service attacks. Such attacks may result in packet loss within wireless networks, consequently inducing collisions. They proposed a solution namely, the RBF sliding mode control method. Numerical examples are applied and it is seen that this method decreases spacing error and maintains a safe framework, but the scope of this study does not cover various cyberattacks.
\par 
Fanid \textit{et. al.}, \cite{Alipour-Fanid2020Impact} worked on the ramifications of jamming attacks and wireless channel fading effects on CACC state space equations to capture their interrelated impacts. Furthermore, they introduced a novel time-infinite main approach to analyze mean string stability. In testing, the jamming attack was initiated beyond the first vehicle following the lead vehicle, progressively advancing upstream within the string. Results indicated that such attacks exerted a more pronounced effect on pushing inter-vehicle distance trajectories towards unsafe states, particularly when the lead vehicle decelerated. Monte Carlo simulations were conducted to assess collision probabilities across the string for different attacker locations. However, this study did not delve into the specifics of inter-vehicle distance trajectories.
Holland \textit{et. al.}, \cite{Holland2024ATesting} developed a testing and verification strategy targeting false data injection, employing particle swarm optimization to fine-tune controller parameters. On average, the optimal solution is achieved after 74 iterations. But, at the implementation time, bounds had to be applied to some parameters, as it damages the hardware.
\par 
Ko \textit{et. al.}, \cite{Ko2021AnApproach} proposed an algorithm named long-short memory-based malicious information detection (LMID). Correlated and non-correlated are the two types of attacks that were considered during the implementation. This algorithm achieves 96\% accuracy when simulated. It helps to achieve string stability from internal attacks. However, various platoon models and trajectories were not considered during the implementation.
\par 
The widespread adoption of ACC necessitates a proactive approach to security. Manufacturers must prioritize robust cybersecurity measures within ACC systems, employing encryption, secure communication protocols, and regular software updates to mitigate potential vulnerabilities.  Furthermore, ongoing collaboration between automotive manufacturers, cybersecurity experts, and regulatory bodies is crucial to establish and enforce rigorous security standards. By prioritizing robust security alongside technological advancements, ensures that ACC continues to enhance driver assistance capabilities without compromising the safety and security of drivers and passengers on the road.
\par 
\subsubsection{V2X}
The evolution of ACC extends beyond maintaining a safe following distance from the car ahead.  Emerging technologies are enabling a new level of communication between ACC systems in different vehicles. This Inter-Vehicle Communication (IVC) unlocks exciting possibilities for enhanced safety and traffic flow.  By exchanging real-time data on speed, position, and intent, ACC systems can collectively optimize traffic flow, anticipate potential hazards beyond the immediate line of sight, and potentially even facilitate cooperative maneuvers.  This introduction of communication protocols into ACC presents both opportunities and challenges that warrant closer examination.
\par 
Liu \textit{et. al.}, \cite{Failure2021ASafety} proposed a safety and secured CACC strategy for Vehicle-to-Vehicle (V2V) communication failure. The control system features a dual-branch control strategy designed to switch to alternative sensors in the event of critical wireless communication failure. Additionally, a linear smooth transition algorithm is integrated to facilitate seamless transitions. An experiment was conducted using platoons consisting of eight vehicles for verification purposes. The findings revealed a reduction in the absolute value of ACC from 3m/s² to approximately 2.3m/s², indicating smooth transitions even in adverse communication conditions. It's important to note that this study does not focus on non-linear smooth transition algorithms.
\par 
Wang \textit{et. al.}, \cite{Wang2017DevelopingPlatoon} proposed is a V2V communication system based on an Eco-CACC framework with the objective of reducing both platoon-wide energy consumption and the emission of hazardous gases. A series of protocols have been devised for various phases of the CACC system, with MATLAB/Simulink employed for simulation in two scenarios: platoon formation and platoon joining along a one-mile segment. Nonetheless, the proposed system demonstrates a reduction in energy consumption by 1.45\% in the former case and 2.17\% in the latter. However, the findings indicate challenges related to braking efficiency and communication latency, which remain unaddressed in this study.
Zhang \cite{Zhang2018Cooperative} proposed a CACC with selective use of wireless V2V communication. The communication system is strained due to the need to monitor the movements of numerous broadcasting vehicles. Moreover, a strategy was outlined to determine whether the received data from other vehicles was utilized by the CACC. Subsequently, a condition is derived for selecting control gains and switching connectivity. To enhance optimization, a data-driven approach is utilized for control gains. Following calculation and simulation, it is observed that the selective CACC enhances safety and reduces uncertainty. However, enhancements in the safety of selective CACC, such as incorporating safety distance constraints and optimizing selection strategies, have not been implemented.
\par 
Stranger \textit{et. al.}, \cite{Stanger2013Model} proposed with a model predictive cooperative ACC approach. They worked on enabling vehicles to interact with their surrounding infrastructure and vehicles to gather information which reduces fuel consumption through predictive vehicle control strategies. Two types of vehicles have been compared and simulated. It turns out that there is a cross benefit of up to 16\% in comparison to ACC-equipped vehicles and up to 20\% concerning uncontrolled predecessors. This approach gives a precise measurement of fuel consumption. However, a further increase in fuel could be achieved with the inclusion of shifting strategies.
\par 
In conclusion, IVC within ACC systems represents a significant leap forward in driver assistance technology. By enabling real-time data exchange between vehicles, ACC with IVC fosters improved safety, smoother traffic flow, and potentially even collaborative maneuvers. However, this advancement necessitates the development of secure and standardized communication protocols. Collaboration amongst automotive manufacturers, regulatory bodies, and cybersecurity experts is crucial to ensure the robustness and reliability of IVC-enabled ACC. As these technologies continue to mature, the future of transportation promises to be not just safer, but also more efficient and collaborative.
\par 
\subsubsection{Energy Conservation}
While ACC is primarily recognized for its safety benefits, its impact extends beyond accident prevention.  Recent studies suggest that ACC can contribute significantly to improved fuel efficiency.  This potential for energy conservation stems from ACC's ability to maintain a consistent speed and minimize unnecessary acceleration and deceleration.  By operating the vehicle within optimal performance parameters, ACC offers a compelling solution for reducing fuel consumption and promoting a more environmentally friendly driving experience.
Guo \textit{et. al.}, \cite{Guo2022Safe} proposed an energy-efficient and secure car-following control approach tailored for intelligent electric vehicles, taking into account regenerative braking. Within the high-level controller, a methodology is introduced to achieve the targeted acceleration. In low-level controllers, tracking and braking control are designed. After testing, it is seen that the approach has a regenerative braking rate of 57.61\%.
\par 
Alrifale \textit{et. al.}, \cite{Alrifaee2015Predictive} formulated a control algorithm based on MPC theory. In this approach, forecasting of traffic patterns and energy optimization practices are computed to increase the vehicle's performance. Simulation results depict 38.38\% and 37.54\% in fuel and energy consumption respectively. But, various factors like the preceding vehicle's velocity, and energy consumption needed to propel the preceding vehicle \textit{etc.} have not been addressed.
He \textit{et. al.}, \cite{He2020TheEnergy} studied the impacts of energy in the ACC. The main motto behind it is the aggravating energy, safety, and environmental issues being faced today. This work mainly focuses on identifying ACC driving behavior and its energy impact with the help of active energy consumption that acts as an energy impact indicator. Regrettably, the findings indicated that followers utilizing ACC contributes to string instability, characterized by the amplification of speed variations downstream. due to their high responsiveness, they dampen when excessive speed increases. And, they use 2.7-20.5\% more energy than human counterparts.
\par 
Liu \textit{et. al.}, \cite{Liu2020Economic} proposed an economic ACC for power split hybrid EV. It aims to improve fuel economy and optimize vehicle route, speed, and power train control. The macroscopic motion planning method optimizes power train control. Whereas, a global power distribution strategy is used to improve the route and speed of the vehicle. A co-simulation model is developed. The findings suggest that the proposed EACC leads to a reduction in fuel consumption of over 30\%. Nonetheless, there is not a perfect alignment between the route, speed, and the power train control strategy.
Wang \textit{et. al.}, \cite{Wang2017Developing} proposed a system for platoon-wide eco-cooperative ACC, primarily aimed at minimizing energy usage. Utilizing MATLAB/Simulink, diverse algorithms are employed for tasks such as sequence determination, gap closing and opening, gap regulation, platoon joining, and splitting. The result showed that the energy consumption was reduced by 1.45\% and 2.17\% in platoon formation and joining respectively. Although most of the uncertainties are handled well, yet certain contingencies such as loss of packets, fading of signals, \textit{etc.,} are not considered.
\par 
Pan \textit{et. al.}, \cite{Pan2023Adaptive} proposed an Economic Adaptive Cruise Control (EACC) incorporating battery aging considerations through Adaptive Model Predictive Control (AMPC). This model comprises two main phases. Initially, it simulates vehicle dynamics, followed by evaluating performance indicators distinguishing driving states. Subsequently, the battery's capacity decay model is developed and enhanced. Experimental results demonstrate the method's effectiveness in optimizing battery life compared to conventional control methods. However, real road condition information and topic-switching methods are not studied in this work.
\par 
In conclusion, ACC emerges as a valuable tool not only for enhanced safety but also for promoting energy conservation. By maintaining a consistent speed and minimizing unnecessary acceleration and deceleration, ACC can demonstrably improve fuel efficiency. This translates to reduced fuel consumption and a smaller environmental footprint. As technology continues to develop, further optimization of ACC's energy-saving capabilities is expected, potentially paving the way for a more sustainable future of transportation.
\par 
\subsubsection{Human Factors}
ACC represents a significant advancement in driver-assistance technology, offering a more relaxed and potentially safer driving experience. However, the effectiveness of ACC hinges not just on sophisticated technology but also on its seamless interaction with the human driver.  This interplay between human factors and the capabilities of ACC warrants closer examination.  Understanding how drivers perceive, interact with, and potentially over-rely on ACC is crucial for optimizing the system's benefits and mitigating potential risks. This analysis of human factors in ACC can guide the development of user-centered interfaces, training programs, and best practices to ensure that ACC truly enhances driving safety and reduces driver fatigue.
Guo \textit{et. al.}, \cite{Guo2020Interacting} contributed to the behavior of human drivers in ACC. 11 participants were asked to volunteer for this study, to generate the driving performance analysis. The results indicate the need for adaptive algorithms majorly in mixed traffic conditions. However, the number of data sources must be increased to obtain accurate results.
\par 
Yao \textit{et. al.}, \cite{Yao2021Target} deliberated on a target vehicle selection algorithm for ACC. The main aim is to improve comfort and safety while the vehicles change lanes. NGSIM dataset was used for the simulation of the proposed work. In the selection algorithm, there are cases namely, safe lane change, cancellation of lane change, and dangerous lane change. Finally, a co-simulation platform is used to check the performance of the algorithm. The outcome of the simulation is that the algorithm guarantees smooth transfer and decreases fluctuation concerning the first two cases. The vehicle's response time significantly increases compared to other algorithms, but functions like the trajectory of the preceding vehicles are not studied.
Althoff \textit{et. al.}, \cite{Althoff2021Provably} presented an approach for provably correct ACC that ensures comfort. They employ a nominal controller safeguarded by a fail-safe controller proven to be correct, minimizing the need for re-certification of the vehicle's safety. A user study indicates that this method enhances user satisfaction, guarantees collision-free operation, and swiftly restores a safe gap following a cut-in event.
\par 
Tajeddin \textit{et. al.}, \cite{Tajeddin2016GMRES} proposed an ecological adaptive cruise control (E-ACC) for plug-in HEV. The E-ACC design integrates principles from generalized minimal residual and non-linear model predictive control, specifically tailored to the Toyota plug-in model. Prioritizing safety and comfort, this model leverages forthcoming trip data and vehicle radar to minimize trip energy costs. Simulation results demonstrate an energy cost enhancement of up to 3.4\%. However, further research could concentrate on refining the controller's energy efficiency even further.
Hu \textit{et. al.}, \cite{Hu2022Trust-Based} proposed a control barrier function approach. The objective is to develop a personalized ACC system founded on individual trust. Three key contributions are presented, firstly, the introduction of a novel quantitative dynamic model delineating driver trust. Secondly, a refined control barrier function methodology is proposed, ensuring system stability. Lastly, a novel prescribed performance function is introduced, eliminating the necessity for precise initial condition values. However, the system does not account for human perceptions of risk and task difficulty.
\par 
Li \textit{et. al.}, \cite{Li2021Improve} contributed to improving ACC under the conditions of driver distraction. Quantitative crash probability models are proposed by considering dynamic traffic situations and driver distraction. After developing and testing this model it is seen that this reduces time headway to a greater extent. However, they consider the speed of the preceding vehicle constant rather than differing it.
Vollrath \textit{et. al.}, \cite{MarkVollrath2011Accident} came up with a driving simulator study. It focuses on the influence of ACC on driving behavior. Besides distance control and beneficial impacts on speed limits, there are signs of delayed reactions in instances demanding immediate braking with cruise control. In order to investigate this proposition, a study was carried out at the German Aerospace Center, involving twenty-two participants navigating various routes under three distinct conditions: ACC, conventional cruise control, and manual driving. The findings revealed that both ACC and CC modes did not incur any violations, yet they also prompted some concerns regarding safety precautions.
\par 
Jiang \textit{et. al.}, \cite{jiang2020APersonalized} proposed a stochastic optimal control algorithm. This algorithm computes the risk of the driver under system uncertainties. After simulation, it is seen that this algorithm displays accurate risk sensitivity, disturbance magnitude \textit{etc.} However, this algorithm is not tested on multiple vehicles at a time.
\par 
In conclusion, achieving optimal safety with ACC necessitates a holistic understanding of human factors. Examining how drivers perceive, interact with, and potentially over-rely on the system is crucial. By addressing these factors through user-centered interface design, targeted training, and clear guidelines, we can ensure responsible use and maximize the benefits of the ACC.  This synergistic approach, combining technological advancements with a nuanced understanding of human behavior, paves the way for a future where ACC empowers drivers, reduces fatigue, and fosters a safer and more efficient transportation landscape.
\par 
\subsubsection{Machine Learning Models}
ACC has revolutionized driver assistance by maintaining a safe following distance. However, the future of ACC lies in its ability to adapt and learn driver behavior. This is where Machine Learning (ML) models come into play. By incorporating ML algorithms, ACC systems can evolve beyond simply reacting to the car ahead.  These models can analyze driver preferences, anticipate upcoming situations, and personalize the ACC experience, potentially leading to a smoother, safer, and more comfortable driving experience.  This integration of machine learning into ACC presents a fascinating new chapter in driver assistance technology.
\par 
Farivar \textit{et. al.}, \cite{Farivar2021SecurityOfNetwork} contributed to the security of the networked control system in ACC. They proposed an artificial neural network identifier to learn the ACC system and predict its operations. Two test scenarios were used to verify the approach. In these scenarios, furtive attacks on ACC were introduced. So they don’t satisfy the speed and space control goals of smart vehicles. The outcomes confirmed that the approach identified the attacks and also mitigated their effects. However, improvements can be made to reduce the false alarm rates.
\par 
Chu \textit{et. al.}, \cite{Chu2021Self-Learning} proposed a self-learning Cruise Control based on an individual car-following system. Initially, a linear quadratic optimal control approach is developed, enabling the derivation of an optimal control law integrating the longitudinal acceleration of the target vehicle. Subsequently, a car-following style learning algorithm is introduced, facilitating the construction of an optimal cruise controller. Upon evaluation, it becomes evident that this controller closely resembles human driver behavior more so than that of ACC. Although it is self-learning, there is still a high probability of accidents as it is closer to human drivers.
Yavas \textit{et. al.}, \cite{Yavas2023Toward} proposed an advance ACC by deep reinforcement learning. It focuses on providing safe and optimal guidelines for maintaining a safe distance while driving behind other vehicles. By simulating the scenarios from real-world driving for evaluating the algorithm, it has been observed to excel in terms of both safety and comfort. However, this algorithm does not have safety monitoring strategies.
\par 
Gao \textit{et. al.}, \cite{Gao2020Personalized} proposed a personalized ACC system that aims at reducing tracking error, acceleration variation, and fuel consumption. This system features three modes representing distinct driving styles, identifiable through a combination of supervised and unsupervised machine learning to determine controller parameters. Test outcomes demonstrate the system's ability to accommodate various driving styles while ensuring both comfort and fuel efficiency. Nevertheless, variations in speed error and distance error are observed across different modes during simulation.
Li \textit{et. al.}, \cite{Li2020Ecological} proposed an EACC to reduce fuel consumption and ensure safety for vehicles with the step-gear transmission. A control strategy employing reinforcement learning utilizing an actor-critic architecture is presented. The controller undergoes evaluation across various driving scenarios, revealing a notable 12.4\% reduction in fuel consumption. Nonetheless, the work does not consider velocity optimization for free-flow urban driving.
\par 
De-Las-Heras \textit{et. al.}, \cite{de2021advanced} developed a prototype ADAS for reading variable messages. This is used to recognize images and indicate locations with percentages. But, the proposed work needs intensive training efforts as it fails to recognize certain static signals. 
Boddupalli \textit{et. al.}, \cite{boddupalli2022resilient} proposed a framework for CACC against V2V attacks named resilient cooperative adaptive cruise control. This is designed to monitor the controller's response time. The proposed machine learning model provides safety while preserving efficiency. However, the work does not focus on detecting sensor attacks.
\par 
In conclusion, the integration of ML models into ACC signifies a pivotal leap forward in driver assistance technology.  By analyzing driver preferences, anticipating road situations, and personalizing the ACC experience, ML models can contribute to a smoother, safer, and more comfortable driving experience.  However, the successful implementation of ML in ACC necessitates ongoing research and development to ensure the robustness and interpretability of these models.  Furthermore, establishing clear ethical guidelines and ensuring driver trust in the technology will be paramount for maximizing the potential of ML-powered ACC.  As this technology matures, we can anticipate a future where ACC seamlessly adapts to individual drivers, creating a more personalized and ultimately safer driving experience.

%% file: includes/tikz-figure/fig-3.tex
\begin{figure*}[h]
	\begin{forest}
		qtree,
		multiple directions={minimum height=4ex, anchor=center, forked edge}
		[ACC%
		[, grow' subtree=east
		[Machine learning models \\ \cite{Farivar2021SecurityOfNetwork} \cite{Chu2021Self-Learning} \cite{Yavas2023Toward} \cite{Gao2020Personalized} \cite{Li2020Ecological} \\ \cite{de2021advanced} \cite{boddupalli2022resilient} \cite{Desjardins2011Cooperative}]
		[V2X \\ \cite{Failure2021ASafety,Wang2017DevelopingPlatoon} \\ \cite{Zhang2018Cooperative} \cite{Stanger2013Model} ]
		[Energy \\conservation\\ \cite{Guo2022Safe} \cite{He2020Multiobjective} \\ \cite{Lu2024Energy} \cite{Alrifaee2015Predictive}%
		[Environmental \\conservation\\ \cite{He2020TheEnergy} \cite{Liu2020Economic} \\ \cite{Wang2017Developing}]
		[Reducing \\battery age\\ \cite{Pan2023Adaptive} \cite{Zlocki2014Methodology} \\ \cite{Li2021Adaptive} \cite{daowd2011Passive}]
		]
		[Human \\ Factors \\ \cite{Larsson2012Driver} \cite{Muslim2019Human} \\ \cite{Fisher2016Human} \cite{Guo2020Interacting}%
		[Driver \\ Behaviour \\ \cite{Yao2021Target} \cite{Althoff2021Provably} \\ \cite{Tajeddin2016GMRES} \cite{Luo2014Adaptive}]
		[Driver \\ Familiarity \\ \cite{Hu2022Trust-Based} \cite{Li2021Improve} \\ \cite{MarkVollrath2011Accident} \cite{jiang2020APersonalized}]
		]		
		]
		[, grow' subtree=west
		[Security\\ \cite{Lee2021Design} \cite{Sawant2021Robust} \\ \cite{Yu2022Safety} \cite{Zhang2020ControlDesign}%
		[Network \\Traffic\\ \cite{Harfouch2018AdaptiveSwitched} \cite{Xing2020Compensation} \\  \cite{Zhang2020ActiveFaultTolerant} \cite{Cui2022Development}]
		[Attacks\\ \cite{Zhang2023Adaptive} \cite{Alipour-Fanid2020Impact} \cite{Yamamoto2021Attack} \cite{Tianxiang2017ACC} \\ \cite{Holland2024ATesting} \cite{Ko2021AnApproach}]
		]
		[Techniques \\ \cite{Zhu2020Synthesis} \cite{Zhu2019LMI} \\ \cite{Junaid2005Intelligent} \cite{Nilsson2014Preliminary} %
		[Collecting data and \\taking the required action\\ \cite{Miyata2010Improvement} \cite{VMilanes2014Modeling} \cite{Hidayatullah2021Adaptive}%
		[Information of the \\preceeding vehicle\\ \cite{Lunze2020Design} \cite{Feng2021Robust} \cite{Brugnolli2019Predictive} \\ \cite{Ma2020Cooperative} \cite{Zhang2021Data-Driven}]
		[Information of all the \\vehicles near it\\ \cite{He2021Defensive} \cite{Razzaghpour2023Predictive}]
		]
		[Collecting previous data and \\predicting the future ones\\ \cite{Pan2022Energy-Optimal} \cite{Wu2019Cooperative}]
		]
		[Safety\\ \cite{Wang2022AFramework} \cite{Chen2021Economic} \cite{Acquarone2023Cooperative} \\ \cite{Yang2020AnAdaptive} \cite{Qin2019Rear-End}%
		[String\\ stability\\ \cite{Gunter2020Model-Based} \cite{Wang2018Cooperative} \cite{Mu2021StringStability} \\ \cite{Tin2019Modeling} \cite{Khattak2023Impact}%
		[Control for\\ stop-and-go\\ \cite{Lu2018NewAdaptive} \cite{Arnaout2014Progressive} \cite{Vahidi2003ResearchAdvances} \cite{Elmorshedy2022Quantitative} \\ \cite{Calvert2020Cooperative} \cite{Milanes2014Cooperative} \cite{Shladover2012Impacts} \cite{CLMelson2018Dynamic}]
		[Spacing\\ \cite{Hu2020Cooperative} \cite{Gunter2021Commercially} \\ \cite{Wang2014AnImproved}]
		[Design\\ \cite{Mutzenich2021Updating} \cite{Yang2021AnOptimization} \\ \cite{Moon2009Design}  \cite{NAUS2010882}]
		]
		[Varying \\factors \\ \cite{Wasserburger2020Probability} \cite{Xiao2023Research} \cite{Shang2021Impacts}%
		[Speed\\  \cite{Wang2021Online} \cite{Mintsis2021Enhanced}  \\ \cite{Diba2014Optimized} \cite{Alankus2020Predictive}]
		[Acceleration\\ \cite{Magdici2017Adaptive} \cite{Pananurak2009ACC} \\ \cite{Woo2019Advanced} ]
		[Road \\conditions\\ \cite{Yang2021Research} \cite{Chen2020Data} \\ \cite{Makridis2020Response} \cite{MANOLIS2020102617} \\ \cite{validi2018Examining}]
		]
		]
		]
		]
	\end{forest}
	\label{fig:Taxonomy}
	\caption{Proposed Classification Taxonomy of the Recent ACC Publications.}	
\end{figure*}
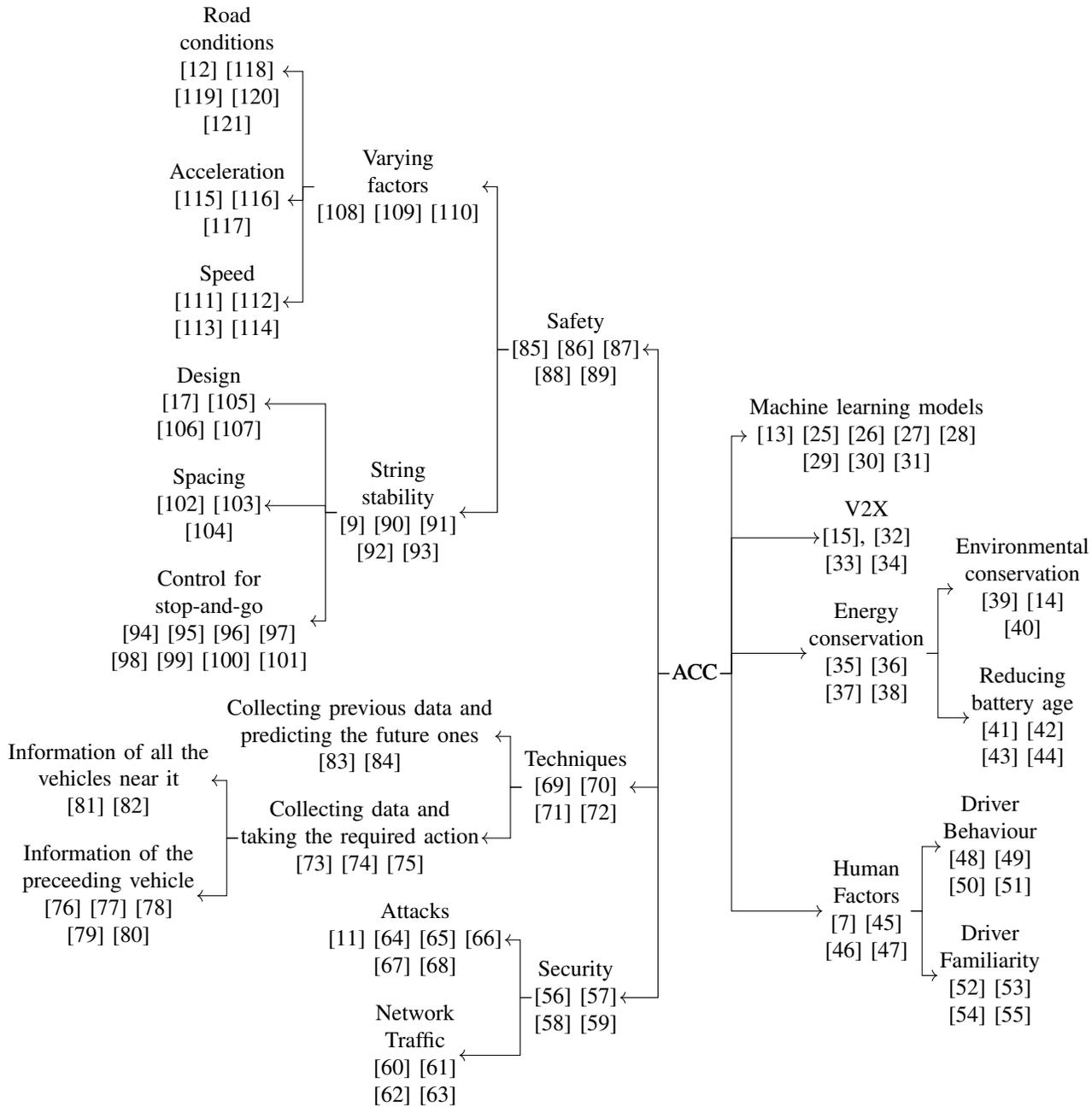

%% file: includes/sections/discussions.tex
\section{Discussions and Insights Gained}
\label{sec:dis}

In previous section we reviewed several works on ACC and utilized them to gain insights and understand the gaps in the design and design. In this section, we provide a tabular comparison of the  works mentioned in the previous section according to our perspective and taxonomy.

\input{includes/tables/sample-lit-comp-table2}

\input{includes/tables/sample-lit-comp-table3}

\input{includes/tables/sample-lit-comp-table4}

\input{includes/tables/sample-lit-comp-table5}

\input{includes/tables/sample-lit-comp-table6}

\input{includes/tables/sample-lit-comp-table7}

\par
Table \ref{tab:table8} delineates various types of ACC and their associated components, factors affecting their performance, and the challenges they encounter. Radar-based systems, reliant on transmitters and receivers, are influenced by environmental conditions such as weather and interference ($C_{1}$, $C_{3}$, $C_{4}$, $C_{5}$, $C_{8}$). Laser-based systems, incorporating LiDAR sensors and optical receivers, confront challenges related to regulatory compliance and safety standards ($C_{1}$, $C_{3}$, $C_{4}$, $C_{5}$, $C_{7}$, $C_{8}$, $C_{9}$, $C_{10}$). Binocular computer vision systems, comprising image sensors and processing units, are constrained by computational resources in accurately interpreting visual data ($C_{1}$, $C_{3}$, $C_{4}$, $C_{5}$, $C_{7}$, $C_{8}$, $C_{9}$, $C_{10}$). Predictive systems, integrating radar and Lidar sensors, play pivotal roles in decision-making and risk management for driving scenarios ($C_{1}$, $C_{3}$, $C_{4}$, $C_{5}$, $C_{7}$, $C_{8}$, $C_{9}$, $C_{10}$). Multi-sensor systems, combining GPS, camera systems, and radar sensors, grapple with the challenge of data processing time for comprehensive situational awareness ($C_{1}$, $C_{3}$, $C_{4}$, $C_{5}$, $C_{8}$). Each system type faces distinct challenges, emphasizing the complexity inherent in developing and implementing ACC technologies ($C_{1}$, $C_{3}$, $C_{4}$, $C_{5}$, $C_{7}$, $C_{8}$, $C_{9}$, $C_{10}$).
\par 

Comparing the parameters outlined in the Table \ref{tab:table2} with summaries of relevant papers, we observe various approaches and their respective advantages and shortcomings in the development of ACC systems. Lin \textit{et. al.} \cite{Lin2020Robust} propose a robust model predictive control strategy to ensure stability in ACC, albeit with limitations on flexibility and applicability. Ko \textit{et. al.} \cite{Ko2021AnApproach} introduce a long-short memory-based malicious information detection method to achieve string stability, yet their approach lacks consideration for diverse platoon models. Holland \textit{et. al.} \cite{Holland2024ATesting} suggest a testing and verification approach to fine-tune controller parameters, though the need for bounds due to hardware damage risks may restrict its utility. Meanwhile, Cui \textit{et. al.} \cite{Cui2022Development} present predecessor-leader following control methods for addressing late responses in ACC, despite facing challenges with high computation time. Alotibi \textit{et. al.} \cite{Alotibi2021Anomaly} propose a kinematic model for anomaly detection in ACC systems, but overlook the impacts of communication delay. Finally, Zhang \textit{et. al.} \cite{Zhang2020ControlDesign} introduce a delay-compensating CACC controller to mitigate communication delay effects, yet their approach does not account for other types of delays. This comparison underscores the complexities involved in designing secure and efficient ACC systems, highlighting trade-offs between stability, response time, and robustness to attacks.
\par
From the comparison between the parameters outlined in the Table \ref{tab:table3} and the summarized approaches from various papers, it becomes evident that different methodologies are employed in the development of ACC systems. Farivar \textit{et. al.} \cite{Farivar2021SecurityOfNetwork} propose an artificial neural network identifier to detect attacks and mitigate their effects, contributing to the security of networked control systems. However, their approach may suffer from reduced improvements in false alarm rates. Chu \textit{et. al.} \cite{Chu2021Self-Learning} introduce a self-learning cruise control system aiming to mimic human drivers' responses, yet this leads to a high probability of accidents despite its ability to adapt to individual car-following styles. Yavas \textit{et. al.} \cite{Yavas2023Toward} present an Advanced ACC powered by deep reinforcement learning, offering safety and comfort benefits, but the absence of safety monitoring systems could be a concern. Similarly, Gao \textit{et. al.} \cite{Gao2020Personalized} proposed a personalized ACC system capable of accommodating different driving styles for fuel conservation, though variations in speed and distance errors across modes may pose challenges. Li \textit{et. al.} \cite{Li2020Ecological} employ a control strategy using reinforcement learning to reduce fuel consumption, but the absence of velocity optimization for free-flow urban driving limits its effectiveness in fuel conservation. Finally, De-Las-Heras \textit{et. al.} \cite{de2021advanced} develop a prototype ADAS capable of recognizing images and indicating location percentages, yet it fails to recognize certain static signals, impacting the interpretation of variable messages. This comparison underscores the diverse strategies and trade-offs inherent in developing effective and reliable ACC systems, emphasizing the importance of addressing key challenges while leveraging advancements in technology for enhanced performance and safety.
\par 
Comparing the summarized approaches from Table \ref{tab:table4} reveals distinct advantages and shortcomings in the development of Adaptive Cruise Control (ACC) systems. Stanger \textit{et. al.} \cite{Stanger2013Model} advocate for Linear Model Predictive Control (MPC), offering fuel consumption measurements while grappling with nonlinearity and model uncertainties. Wang \textit{et.  al.} \cite{Wang2017DevelopingPlatoon} propose a V2V-based Eco-CACC system, effectively reducing energy consumption and emissions, although challenges persist regarding braking ability and communication delays. Liu \textit{et. al.} \cite{Liu2020Economic} present an Extended ACC strategy utilizing genetic algorithms to achieve reduced fuel consumption, yet overlook discussions on V2V and V2I communication with adhoc networks, limiting applicability to Hybrid Electric Vehicles (HEV). Conversely, Zhang \cite{Zhang2018Cooperative} investigates a CACC design selectively incorporating motion data, enhancing safety and reducing perturbations, despite the absence of optimization strategies for selection. This comparison underscores the need for comprehensive approaches addressing communication challenges and optimizing system performance to enhance safety and efficiency in ACC systems.
\par
Comparing the parameters outlined in the Table \ref{tab:table5} with the summaries of relevant papers, it is evident that every approach offers specific advantages and shortcomings in the field of energy conservation in Adaptive Cruise Control (ACC) systems. He \textit{et. al.} \cite{He2020TheEnergy} focus on tractive energy consumption, ensuring string stability and potential for performance evaluation through test track studies. Similarly, Liu \textit{et. al.} \cite{Liu2020Economic} propose macroscopic motion planning, focusing on fuel consumption reduction, despite misalignment issues between route and speed with powertrain control strategies. Wang \textit{et. al.} \cite{Wang2017Developing} introduce a platoon-wide Eco-Cooperative ACC system aimed at minimizing energy consumption, yet they overlook contingencies such as packet loss and signal fading, potentially affecting system reliability. In contrast, Pan \textit{et. al.} \cite{Pan2023Adaptive} present an economic adaptive cruise controller, extending battery service life while enhancing road realism and control under diverse conditions. Guo \textit{et. al.} \cite{Guo2020Safe} advocate for an AFSMC control method, enabling breaking energy recovery without compromising safety, though their assumption of vehicles driving on flat roads during simulations may limit real-world applicability. Finally, Alrifale \textit{et. al.} \cite{Alrifaee2015Predictive} propose a control algorithm based on MPC theory to reduce fuel consumption, but fails to identify fuel conservation during unexpected scenarios. This comparison underscores the importance of considering various factors such as system reliability, adaptability to real-world conditions, and comprehensive energy conservation strategies in the development of ACC systems.
\par 
The table \ref{tab:table6} presents a range of techniques proposed in research publications aimed at improving various aspects of automotive systems. Zhu \textit{et. al.} \cite{Zhu2020Synthesis} introduced control synthesis for maintaining string stability, although limited to homogeneous platoons. Wu \textit{et. al.} \cite{Wu2019Cooperative} presented an algorithm with a Kalman filter enhancing communication loss resilience but lacked discussion on platoon stability. Miyata \textit{et. al.} \cite{Miyata2010Improvement} developed an ACC system considering vehicle slip side, reducing driver burden but without mentioned performance optimization schemes. Feng \textit{et. al.} \cite{Feng2021Robust} introduced a gain scheduling technique ensuring robustness under varying mass conditions but didn't focus on string stability. Brugnolli \textit{et. al.} \cite{Brugnolli2019Predictive} implemented Tube Model Predictive Control handling uncertainty with less computational burden but didn't discuss efficiency. Ma \textit{et. al.} \cite{Ma2020Cooperative} proposed a model predictive control technique successfully tracking changing speeds but overlooked controller efficiency. He \textit{et. al.} \cite{He2021Defensive} introduced MPC based on a swarm optimization algorithm, maintaining better distancing with quick response but causing little vibration in acceleration. Lunze et al. (2021) proposed a Defensive Ecological ACC algorithm reducing dwelling time in blind spot zones, yet its performance under uncertainty remains unmeasured. Hidayatullah \textit{et. al.} \cite{Hidayatullah2021Adaptive} suggested local vehicle controllers and communication structure resolving communication disputes but unsolved feedback controller design remains a potential barrier. These techniques offer advancements in control, communication, and predictive modeling, providing benefits such as stability, resilience, and efficiency, yet they also face challenges such as limited applicability, efficiency concerns, and unresolved design barriers.

\par 
The table \ref{tab:table7} outlines evaluations of research publications based on safety in automotive systems. Control synthesis methods such as those proposed by Lin \textit{et. al.} \cite{Lin2020Robust} maintain string stability, though limited to homogeneous platoons. Model-based approaches like the one discussed by Farivar \textit{et. al.} \cite{Farivar2021SecurityOfNetwork} focusing on string stability using field data offer improvements but lack discussion on commercial ACC enhancements. Personalized ACC systems, as introduced by Lin \textit{et. al.} \cite{Lin2020Robust}, cater to different driving styles, providing comfort, yet they encounter errors in speed and distance under various modes. Probability-based velocity prediction methods, like the one described in the study by Yao \textit{et. al.} \cite{Yao2021Target}, save energy with accurate predictions but focus solely on velocity. Smart Driver Models, exemplified by Lu \textit{et. al.} \cite{Lu2018NewAdaptive}, offer linear stability, yet they overlook lane-changing behavior. Neural network identifiers for attack detection, as explored by Farivar \textit{et. al.} \cite{Farivar2021SecurityOfNetwork}, offer robust defense but does not entirely prevent attacks. These studies collectively contribute to safety advancements in automotive systems, offering solutions for stability, personalized driving experiences, and attack detection, yet they may encounter limitations in addressing heterogeneous scenarios, human factors, and comprehensive attack prevention.
\par

 From the review of the literature we were able to identify few gaps in the research which are mentioned below .
\par 
%
\begin{itemize}
	\item The previous studies have heavily relied on simulations which makes it difficult for implementations in real-life basis as it lacks in actual data to validate and asses the proposed models reliability.
	 
	\item Many of the researchers have assumed the speed of the lead vehicle as constant, while proposing their ACC system. This neglects the dynamics nature of the real world traffic, thus effecting the overall accuracy of these systems. 
	
	\item In scenarios where an ACC-equipped vehicle encounters a non-ACC equipped vehicle, conflicts arise due to the differing driver behavior. The ACC-equipped vehicle faces challenges due to the unpredictability of the manually driven vehicles. 

	\item The ml-based ACC systems tend to achieve higher accuracy but, one notable concerns are its high false-alarm rates which can be a crucial threat in real-life situations.
	
	\item The considerations of the human behavioral inputs falls shorts in most of the literature. The Human FRTD plays an major role role in recognizing the functionalities required by the ACC to sustain in the real world.
	
	\item Many of the present work focus on a small set of factors while designing their ACC systems, which limits the reliability of the ACC in cases where it encounters unexpected scenarios. Therefore, a comprehensive design for the ACC is required which considers most of the affecting factors. 
\end{itemize}

%% file: includes/tables/sample-lit-comp-table2.tex
\begin{table*}
	\centering
	\caption{Evaluation of Reviewed Research Publications: Machine Learning Models.}
    \label{tab:table2}
	\ra{1.8}
	\begin{tabular}{@{}
				m{0.05\textwidth}
				m{0.05\textwidth}
				m{0.1\textwidth}
				m{0.1\textwidth}
				m{0.1\textwidth}
				m{0.1\textwidth}
				m{0.05\textwidth}
				m{0.04\textwidth}
			    m{0.04\textwidth}
			    m{0.04\textwidth}
	}
	\midrule 
	\multicolumn{1}{c}{\textbf{Author}} &
	\multicolumn{1}{c}{\textbf{Year}} & 
	\multicolumn{1}{c}{\textbf{Approach}} & 
	\multicolumn{1}{c}{\textbf{Advantages}} & 
	\multicolumn{1}{c}{\textbf{Shortcomings}} & 
	\multicolumn{1}{c}{\textbf{\makecell{Cause of \\ usage}}} &
	\multicolumn{1}{c}{\textbf{\makecell{Model \\ complexity}}} &
	\multicolumn{1}{c}{\textbf{\makecell{Handles \\ imbalanced \\ data}}} &
	\multicolumn{1}{c}{\textbf{\makecell{Handles \\  missing \\ data}}} &
	\multicolumn{1}{c}{\textbf{\makecell{Handeling \\ large \\ data}}} \\	
			
	\toprule
 
     Farivar \textit{et.al.}, \cite{Farivar2021SecurityOfNetwork} & 2021 & artificial neural network identifier & Identify attacks and mitigate their effects & reduction of improvements on false alarm rates & security of networked control system & \checkmark & \checkmark & $\times$ & \checkmark\\  
     
     Chu \textit{et.al.}, \cite{Chu2021Self-Learning} & 2021 & self-learning CC & controller's response is close to that of human drivers & high probability of accidents & automatically adapt to individual car-following style & \checkmark & $\times$ & $\times$ & \checkmark\\
     
     Yavas \textit{et.al.}, \cite{Yavas2023Toward} & 2023 & Advanced ACC powered by DRL & provides safety and comfort & does not have safety monitoring systems & provide safe and comfortable car-following policies & \checkmark & $\times$ & $\times$ & \checkmark\\
     
     Gao \textit{et.al.}, \cite{Gao2020Personalized} & 2020 & personalized ACC system & system can meet different driving styles with fuel conversation & speed error and distance error are different in different modes & reduce tracking error and fuel consumption & \checkmark & $\times$ & $\times$ & \checkmark\\
    
     Li \textit{et.al.}, \cite{Li2020Ecological} & 2020 & control strategy using reinforcement learning & reduction in fuel consumption & absence of velocity optimization for free-flow urban driving & fuel conservation & $\times$ & $\times$ & \checkmark & \checkmark\\
     
     De-Las-Heras \textit{et.al.}, \cite{de2021advanced} & 2021 & prototype ADAS & recognizes images and indicate location with percentages & fails to recognize certain static signals & reading variable messages & \checkmark& \checkmark & \checkmark & \checkmark  \\
     
     Boddupalli \textit{et.al.}, \cite{boddupalli2022resilient} & 2022 & framework for CACC & provides safety & Does not detect sensor attacks & focuses on reducing V2V attacks & \checkmark & $\times$ & $\times$ & \checkmark\\

    \bottomrule
	\end{tabular} 	
\end{table*}

%% file: includes/tables/sample-lit-comp-table3.tex
\begin{table*}
	\centering
	\caption{Evaluation of Reviewed Research Publications: V2X.}
	\label{tab:table3}
	\ra{1.8}
	\begin{tabular}{@{}
				m{0.1\textwidth}
				m{0.05\textwidth}
				m{0.1\textwidth}
				m{0.1\textwidth}
				m{0.1\textwidth}
				m{0.15\textwidth}
				m{0.05\textwidth}
				m{0.05\textwidth}
				m{0.05\textwidth}
				m{0.05\textwidth}
	}
	\midrule 
	\multicolumn{1}{c}{\textbf{Author}} &
	\multicolumn{1}{c}{\textbf{Year}} & 
	\multicolumn{1}{c}{\textbf{Approach}} & 
	\multicolumn{1}{c}{\textbf{Advantages}} & 
	\multicolumn{1}{c}{\textbf{Shortcomings}} & 
	\multicolumn{1}{c}{\textbf{Type of communication}} &
	\multicolumn{1}{c}{\textbf{QoS}} &
	\multicolumn{1}{c}{\textbf{Reliable}} & 
	\multicolumn{1}{c}{\textbf{\makecell{Error \\ handling}}} & 
	\multicolumn{1}{c}{\textbf{Coverage}} \\
			
	\toprule 
			
    Stanger \textit{et.al.} \cite{Stanger2013Model} & 2013 & Linear model predictive control (MPC) & Precise measurements of fuel consumption & Linear MPC cannot handle non linearity and model uncertainties & Integration V2V and V2I communication with dynamics of the system & \checkmark & \checkmark &\checkmark & \checkmark\\
		    
	Wang \textit{et.al.}, \cite{Wang2017DevelopingPlatoon} & 2017 & V2V communication based on Eco-CACC system & reduces energy consumption and  emission of hazardous gases & issues of braking ability and communication delay is not addressed & reduces platoon wide energy consumption in V2V & \checkmark & \checkmark &\checkmark & \checkmark\\    
		    
    Liu \textit{et.al.} \cite{Failure2021ASafety} & 2020 & Extended ACC strategy based on generic algorithm & Reduces fuel consumption & V2V and V2I communication with adhoc network is not discussed & Hybrid Electric Vehicle (HEV) & $\times$ & \checkmark & \checkmark & $\times$\\
		    
	Zhang \cite{Zhang2018Cooperative} & 2018 & Investigated design of CACC that selectively incorporated motion data of other vehicles. & Improves safety and reduces perturbations & Optimization of selection strategies are not investigated. & Imparting with V2V in mixed traffic. & \checkmark & \checkmark &\checkmark & \checkmark\\
			
	\bottomrule
	\end{tabular} 	
\end{table*}

%% file: includes/tables/sample-lit-comp-table4.tex
\begin{table*}
	\centering
	\caption{Evaluation of Reviewed Research Publications: Energy Conservation.}
	\label{tab:table4}
	\ra{1.8}
	\begin{tabular}{@{}
				m{0.1\textwidth}
				m{0.05\textwidth}
				m{0.1\textwidth}
				m{0.1\textwidth}
				m{0.15\textwidth}
				m{0.05\textwidth}
				m{0.05\textwidth}
				m{0.05\textwidth}
				m{0.05\textwidth}
	}
	\midrule 
	\multicolumn{1}{c}{\textbf{Author}} &
	\multicolumn{1}{c}{\textbf{Year}} & 
	\multicolumn{1}{c}{\textbf{Approach}} & 
	\multicolumn{1}{c}{\textbf{Advantages}} & 
	\multicolumn{1}{c}{\textbf{Shortcomings}} & 
	\multicolumn{1}{c}{\textbf{\makecell{Energy \\ conservation}}} & 
	\multicolumn{1}{c}{\textbf{\makecell{Battery \\ age}}} &
	\multicolumn{1}{c}{\textbf{\makecell{Regenrative \\ breaking}}} &
	\multicolumn{1}{c}{\textbf{\makecell{Fuel efficiency}}} \\
			
	\toprule
		
	He \textit{et.al.} \cite{He2020TheEnergy} & 2020 & Tractive energy consumption & String stability & Study on test tracks could be performed. & \checkmark & $\times$ & \checkmark & \checkmark\\

    Liu \textit{et.al.}, \cite{Liu2020Economic} & 2020 & Macroscopic motion planning & reduces fuel consumption & route and speed are not perfectly aligned with power train control strategy & \checkmark & $\times$ & \checkmark & \checkmark\\
			
	Wang \textit{et.al.}, \cite{Wang2017Developing} & 2017 & platoon-wide Eco-Cooperative Adaptive Cruise Control system & minimizing energy consumption & certain contingencies such as loss of packets, fading of signals, \textit{etc.} are not considered. & \checkmark & $\times$ & $\times$ & $\times$ \\  	
		
	Pan \textit{et.al.} \cite{Pan2023Adaptive} & 2023 & Economic adaptive cruise controller & Extends the service life of battery & Can increase road realism and optimal control under different working conditions. & $\times$ & \checkmark & \checkmark & \checkmark\\
			
	Guo \textit{et.al.} \cite{Guo2022Safe} & 2022 & AFSMC control method & Breaking energy recovery with no loss of safety & During simulation it is assumed that vehicles drive in flat road. & \checkmark & $\times$ & \checkmark & \checkmark\\
	
	Alrifale \textit{et.al.}, \cite{Alrifaee2015Predictive} & 2015 & control algorithm based on MPC theory & reducing fuel consumption & conservation of fuel during unexpected scenarios not addressed & \checkmark & $\times$ & \checkmark & \checkmark \\
	 
%
			
	\bottomrule
	\end{tabular} 	
\end{table*}

%% file: includes/tables/sample-lit-comp-table5.tex
\begin{table*}
	\centering
	\caption{Evaluation of Reviewed Research Publications: Techniques.}
	\label{tab:table5}
	\ra{1.8}
	\begin{tabular}{@{}
				m{0.1\textwidth}
				m{0.05\textwidth}
				m{0.1\textwidth}
				m{0.15\textwidth}
				m{0.1\textwidth}
				m{0.05\textwidth}
				m{0.05\textwidth}
				m{0.05\textwidth}
				m{0.05\textwidth}
	}
	\midrule 
	\multicolumn{1}{c}{\textbf{Author}} &
	\multicolumn{1}{c}{\textbf{Year}} & 
	\multicolumn{1}{c}{\textbf{Approach}} & 
	\multicolumn{1}{c}{\textbf{Advantages}} & 
	\multicolumn{1}{c}{\textbf{Shortcomings}} &  
	\multicolumn{1}{c}{\textbf{Prediction}} &
	\multicolumn{1}{c}{\textbf{\makecell{Collecting\\data}}} &
	\multicolumn{1}{c}{\textbf{\makecell{Emergency \\ preparedness}}} &
	\multicolumn{1}{c}{\textbf{Risk assessment}} \\
			
	\toprule
			
	Zhu \textit{et.al.} \cite{Zhu2020Synthesis} & 2020 & Control synthesis & Maintains string stability & Limited to homogeneous platoon & \checkmark & $\times$ & \checkmark & \checkmark \\
			
	Jia \textit{et.al.} \cite{Pan2022Energy-Optimal} & 2020 & Receding horizon dynamic programming solution (RH-DP) & Better than Linear Model Predictive Control (LMPC) & Didn't focus on energy effeciency & $\times$ & \checkmark & \checkmark & \checkmark\\
	
	Wu \textit{et.al.}, \cite{Woo2019Advanced} & 2019 & algorithm with a Kalman filter &  The information is fed to the ego-vehicle CACC controller in case of communication loss & issues like platoon stability is not discussed & $\times$ & \checkmark & \checkmark & \checkmark \\
	
	Miyata \textit{et.al.}, \cite{Miyata2010Improvement} & 2010 & ACC system with the following control that considers the vehicle’s slip side. & system reduces the burden on the driver & performance optimization schemes are not mentioned & $\times$ & \checkmark & \checkmark & \checkmark \\
			
	Hidayatullah \textit{et.al.} \cite{Hidayatullah2021Adaptive} & 2021 & Gain scheduling technique & Robustness is maintained even after varying the mass & Didn't focus on string stability with varying mass & $\times$ & \checkmark & \checkmark & \checkmark\\

    Feng \textit{et.al.}, \cite{Feng2021Robust} & 2021 & Tube Model Predictive control & handles uncertainty with less burden of communication and computation & Advancements in integration of prediction models & $\times$ & \checkmark & \checkmark & \checkmark\\

    Brugnolli \textit{et.al.}, \cite{Brugnolli2019Predictive} & 2019 & model predictive control technique & controller satisfactorily tracked the changing speeds and maintained a safe distance from the preceding vehicle &  efficincy of the controller performance is not discussed & $\times$ & \checkmark & \checkmark & \checkmark\\

    Ma \textit{et.al.}, \cite{Ma2020Cooperative} & 2020 & MPC based on swarm optimization (SA-PSO) algorithm & It maintains better distancing, reduces spacing and has quick response with time delay & little vibration in acceleration & $\times$ & \checkmark & $\times$ & \checkmark\\
    
	He \textit{et.al.} \cite{He2021Defensive} & 2021 & Defensive ecological ACC (DEco-ACC) algorithm & reduce dwelling time in neighboring vehicles' blind spot zones (BSZs) & DEco-ACC's performce isn't measured when it is uncertain & $\times$ & \checkmark & \checkmark & \checkmark \\
			
	Lunze \textit{et.al.} \cite{Lunze2020Design} & 2020 & Local vehicle controllers and communication structure & By solving inequality the communication dispute can be resolved & Design of feedback controller could be a barrier which is unsolved &  $\times$ & \checkmark & \checkmark & \checkmark\\
			
	\bottomrule
	\end{tabular} 	
\end{table*}

%% file: includes/tables/sample-lit-comp-table6.tex
\begin{table*}
	\centering
	\caption{Evaluation of Reviewed Research Publications: Safety.}
	\label{tab:table6}
	\ra{1.8}
	\begin{tabular}{@{}
				m{0.1\textwidth}
				m{0.1\textwidth}
				m{0.1\textwidth}
				m{0.1\textwidth}
				m{0.15\textwidth}
				m{0.05\textwidth}
				m{0.05\textwidth}
				m{0.05\textwidth}
				m{0.05\textwidth}
	}
	\midrule 
	\multicolumn{1}{c}{\textbf{Author}} &
	\multicolumn{1}{c}{\textbf{Year}} & 
	\multicolumn{1}{c}{\textbf{Approach}} & 
	\multicolumn{1}{c}{\textbf{Advantages}} & 
	\multicolumn{1}{c}{\textbf{Shortcomings}} & 
	\multicolumn{1}{c}{\textbf{\makecell{Various \\ factors}}} & 
	\multicolumn{1}{c}{\textbf{Stability}} &
	\multicolumn{1}{c}{\textbf{\makecell{Gradual \\ adjustemnts}}} &
	\multicolumn{1}{c}{\textbf{\makecell{Safety \\ override}}} \\
			
	\toprule
			
	Zhu \textit{et.al.} \cite{Zhu2020Synthesis} & 2020 & Control synthesis & Maintains string stability & Limited to homogeneous platoon & \checkmark & $\times$ & \checkmark & \checkmark  \\
			
	Gunter \textit{et.al.}, \cite{Gunter2020Model-Based} & 2020 & model-based string stability of ACC system using field data & string stability & improvising the commercials ACC were not discussed & $\times$ & \checkmark & \checkmark & $\times$ \\  
			
	Yang \textit{et.al.}\cite{Yang2021Research} & 2021 & ACC system architecture & Good applicability to road conditions & There is a slight overshoot in the result when testing front car cut-in and cut-out & \checkmark & $\times$ & \checkmark & $\times$\\
		
	Gao \textit{et.al.} \cite{Gao2020Personalized} & 2020 & Personalized ACC based system on driving style recognition & Can meet different driving styles with guarenteed comfort & Speed and distance error are differen under different modes & \checkmark & $\times$ & \checkmark & \checkmark \\
			
	Wasserburger \textit{et.al.} \cite{Wasserburger2020Probability} & 2020 & Probability-based short-term velocity prediction method & Saves more energy, accurate velocity prediction & Focused only on velocity & \checkmark & $\times$ & \checkmark & $\times$\\
			
	Yao \textit{et.al.} \cite{Yao2021Target} & 2021 & Target vehicle selection algorithm & Respond to lane chnage of preceding vehicle in advance & Did not focus on trajectories of preceding vehicle  & \checkmark & $\times$ & \checkmark & $\times$ \\
			
	Hu \textit{et.al.} \cite{Hu2022Trust-Based} & 2022 & Control barrier function & Guarantes stability & Can focus on capturing human's FRTD status & $\times$ & \checkmark & \checkmark & \checkmark\\
		
	Lu  \textit{et.al.} \cite{Lu2018NewAdaptive} & 2018 & Smart Driver Model (SDM) & Linear stability & Lane-changing behaviour is ignored & $\times$ & \checkmark & \checkmark & $\times$\\
			
	Lin \textit{et.al.} \cite{Lin2020Robust} & 2020 & Robust model predictive control (RMPC) & predicts sequence of disturbance & Lyapunor function restricts the flexibility of RMPC & $\times$ & \checkmark & \checkmark & $\times$\\
			
	Farivar \textit{et.al.} \cite{Farivar2021SecurityOfNetwork} & 2021 & Neytral network identifier & detects covert attacks & Can block the attacks instead of mitigating it & $\times$ & \checkmark & \checkmark & $\times$ \\ 
			
	\bottomrule
	\end{tabular} 	
\end{table*}

%% file: includes/tables/sample-lit-comp-table7.tex
\begin{table*}[!h]
	\centering
	\caption{Evaluation of Reviewed Research Publications: Human Factors}
	\label{tab:table7}
	\ra{1.8}
	\begin{tabular}{@{}
			m{0.1\textwidth}
			m{0.1\textwidth}
			m{0.1\textwidth}
			m{0.1\textwidth}
			m{0.1\textwidth}
			m{0.05\textwidth}
			m{0.05\textwidth}
			m{0.05\textwidth}
			m{0.05\textwidth}
		}
		\midrule 
		\multicolumn{1}{c}{\textbf{Author}} &
		\multicolumn{1}{c}{\textbf{Year}} & 
		\multicolumn{1}{c}{\textbf{Approach}} & 
		\multicolumn{1}{c}{\textbf{Advantages}} & 
		\multicolumn{1}{c}{\textbf{Shortcomings}} & 
		\multicolumn{1}{c}{\textbf{Familiarity}} & 
		\multicolumn{1}{c}{\textbf{Behavior}} &
		\multicolumn{1}{c}{\textbf{UID}} &
		\multicolumn{1}{c}{\textbf{Feedback mechanism}}\\
		
		\toprule
		
		Yao \textit{et.al.} \cite{Yao2021Target} & 2021 & target vehicle selection algorithm & improves comfort and safety while the vehicles change lanes & functions like the trajectory of the preceding vehicles are not studied & $\times$ & \checkmark & $\times$ & \checkmark\\
	
		Althoff \textit{et.al.} \cite{Althoff2021Provably} & 2021 & provably-correct adaptive cruise controller & ensures comfort, even in the event of cut-ins & approach does not impede user satisfaction & $\times$ & \checkmark & $\times$ & \checkmark\\
		
		Tajeddin \textit{et.al.} \cite{Tajeddin2016GMRES} & 2016 & Non linear model predictive control & uses inforation to maintain the safety of the driver & the predictions are not accurate & $\times$ & \checkmark & $\times$ & $\times$ \\
		
		Luo \textit{et.al.} \cite{Luo2014Adaptive} & 2014 & ACC control algorithm & provides driving safety and comfort & string stability is not addressed & $\times$ & \checkmark & \checkmark & \checkmark\\
		
		Hu \textit{et.al.} \cite{Hu2022Trust-Based} & 2022 & dynamic model describing the driver’s trust on the ACC longitudinal driving & guarantee stability and satisfy system constraints & FRTD status is not considered & \checkmark & $\times$ & $\times$ & \checkmark\\
	
		Li \textit{et.al.} \cite{Li2021Improve} & 2021 & time headway model & compensate for the extra crash risk while being distracted & various factors like speed of the vehicle are not considered & \checkmark & $\times$ & $\times$ & \checkmark \\
		
		\bottomrule
	\end{tabular} 	
\end{table*}

%% file: includes/sections/future.tex
\section{Future Research Directions}
\label{sec:future}

The work done by recent studies on ACC and ADAS is promising, but further research is needed to optimize safety measures for diverse driving conditions. The study on ADS- and ADAS-involved crashes highlights the dominance of rear-end collisions and the influence of factors such as vehicle speed and road type, suggesting a need for improved control algorithms and human-machine interface designs \cite{ADAS12024}. Additionally, the development of a spatio-temporal graph transformer-based prediction framework for evasive behavior and collision risk shows significant advancements in driver satisfaction and safety under near-crash scenarios (\cite{ADAS22024}). Furthermore, the analysis of AEB systems underscores the importance of understanding ADAS sensor functionality and operating algorithms to enhance accident investigation and forensic analysis \cite{ADAS32024}. Future research should focus on integrating these predictive models with real-time data to enhance the responsiveness and reliability of ADAS and ADS systems, ultimately improving traffic safety in mixed traffic environments. In this section, we provide recommendations for the identified gaps helps the researchers to carry on advancements in the field of cruise control to bring out a stable automated transportation.

\begin{itemize}
	\item \textit{Considerations of real world aspects} - Integration of the real world variabilities is an effective factor for enhancing the accuracies of the ACC systems, presently the models are evaluated in simulated platforms by considering a few of the real world aspects. The researchers need to integrate real world data by deploying sensors like radars, liDARs, and cameras to capture data that reflects real-world diversity. This can ultimately lead to the development of more optimized and robust ACC systems that prove to be reliable in life-threatening situations.
	
	\item \textit{User Inputs} - consideration of human behavioral parameters is one thing which the present day ACC systems lack in, by allowing the users to personalize the ACC by allowing to set their parameters such as distance from preceding vehicle, acceleration, \textit{etc.} Further research is to be made on integrating FRTD parameters with the addition of customization for the drivers during the decision making process. This can be done by allowing the usage of hand gestures, eye movements, facial expression, \textit{etc.} Therefore integrating these user-centric preferences can help faster adoption of the ACC systems.
	
	\item \textit{Use of ML} - The use of machine learning techniques promises prominent future directions of the advancement in ACC. This helps to improve the performance of the ACC systems to predict and anticipate changes in traffic patterns and driving conditions. The one associated problem is the high false alarm rates which are to be dealt with by providing the ML-model high quality real life data and by leveraging feature engineering. Thus, ML techniques provide a powerful direction to advance the ACC technology.
	
	\item \textit{Social Relations} - The use of social relations as a future direction can help in enhancing the ACC systems. By integrating advanced V2V communications, the vehicles can cooperate to optimize the traffic flow, reduce congestion, \textit{etc.} Further, the ACC systems can be improvised by making them aware of other road entities like pedestrians and cyclists, \textit{etc.,} that may come in contact with the vehicles. Therefore, it is needed to incorporate ethical decision-making practices into the ACC systems. This can help in anticipating and adapting to the dynamic interactions in daily life.
 	
\end{itemize}	
\input{includes/tables/sample-lit-comp-table9}
Table \ref{tab:table9} presents a comprehensive overview of challenges encountered in the development and implementation of ACC, along with insights from existing literature, identified gaps, and future directions for addressing these challenges. C1 highlights the need for improvement in control performance through AI algorithms for better real-time awareness in autonomous driving. C2 addresses poor road conditions exacerbated by adverse weather challenges, advocating for the development of resilient infrastructure to mitigate their effects. C3 emphasizes the necessity for further research to enhance string stability in commercial ACC systems through adaptive control algorithms. C4 focuses on improving object detection algorithms to address inaccuracies in object detection. C5 underscores the importance of enhancing sensor durability to ensure stable strings in vehicle safety. C6 suggests enhancing driver training and system feedback to promote effective velocity optimization in ACC, aiming to address potential driver misuse. C7 advocates for innovation in model-based control strategies to improve theoretical work on string stability. C8 proposes AI sensor optimization to mitigate the extensive manual work and high costs associated with time-of-flight systems. C9 highlights the importance of prioritizing comprehensive real-life testing to ensure regulatory compliance. Finally, C10 stresses the need for developing robust communication protocols to address communication loss on the road. These insights, drawn from a variety of sources, offer a roadmap for tackling the challenges and advancing the field of ACC technology.
\par 

%% file: includes/tables/sample-lit-comp-table9.tex
\begin{table*}
	\centering
	\caption{Insights}
	\label{tab:table9}
	\ra{1.8}
	\begin{tabular}{@{}
			m{0.05\textwidth}
			m{0.2\textwidth}
			m{0.2\textwidth}
			m{0.2\textwidth}
		}
		\midrule 
		\multicolumn{1}{c}{\textbf{Challenge}} &
		\multicolumn{1}{c}{\textbf{Existing Literature}} & 
		\multicolumn{1}{c}{\textbf{Gap}} & 
		\multicolumn{1}{c}{\textbf{Future direction}} \\ 
		
		\toprule
		$C_{1}$ & \cite{Chen2021Economic}, \cite{Vahidi2003ResearchAdvances}, \cite{Wang2018Cooperative}, \cite{Elmorshedy2022Quantitative}, \cite{Moon2009Design}, \cite{Chen2020Data}, \cite{Makridis2020Response} & Improvement in control performance & AI algorithms for better real-time awareness in autonomous driving.\\
		
		$C_{2}$ & \cite{Pan2023Adaptive}, \cite{Wang2017Developing}, \cite{Liu2020Economic}, \cite{Guo2022Safe}, \cite{Stanger2013Model}, \cite{Failure2021ASafety}, \cite{Ko2021AnApproach} & Poor road conditions are linked to adverse weather challenges & develop resilient road infrastructure caused by adverse weather\\
	
		$C_{3}$ & \cite{Gunter2020Model-Based}, \cite{Lu2018NewAdaptive}, \cite{Arnaout2014Progressive}, \cite{Calvert2020Cooperative}, \cite{Milanes2014Cooperative}, \cite{CLMelson2018Dynamic}, \cite{Mintsis2021Enhanced} & String instability in commercial ACC systems necessitates further research for improvement & Enhance string stability through adaptive control algorithms.\\
		
		$C_{4}$ & \cite{Farivar2021SecurityOfNetwork}, \cite{Yao2021Target}, \cite{Alrifaee2015Predictive}, \cite{Zhang2018Cooperative}, \cite{Wang2017DevelopingPlatoon}, \cite{Cui2022Development}, \cite{Zhang2020ActiveFaultTolerant} & inaccuracy in addressing object detection & Improve object detection algorithms\\
		
		$C_{5}$ & \cite{Gunter2021Commercially}, \cite{Wang2021Online}, \cite{MANOLIS2020102617}, \cite{Pan2022Energy-Optimal}, \cite{Hidayatullah2021Adaptive}, \cite{He2021Defensive}, \cite{Harfouch2018AdaptiveSwitched} & Unstable string emphasizes the need for robust sensors in vehicle safety. & Enhance sensor durability for stable strings\\ 
		
		$C_{6}$ & \cite{Li2020Ecological}, \cite{Yavas2023Toward}, \cite{jiang2020APersonalized}, \cite{Li2021Improve}, \cite{Hu2022Trust-Based}, \cite{Guo2020Interacting}, \cite{He2020TheEnergy} & Neglecting velocity optimization hints at potential driver misuse in ACC & Enhance driver training and system feedback to promote effective velocity optimization in ACC\\ 
		
		$C_{7}$ & \cite{Mu2021StringStability}, \cite{Hu2020Cooperative} \cite{Shladover2012Impacts}, \cite{Wang2014AnImproved}, \cite{Magdici2017Adaptive}, \cite{Miyata2010Improvement}, \cite{Lunze2020Design}, \cite{Feng2021Robust}, \cite{Brugnolli2019Predictive} & Theoretical work on string stability & Innovation of model-based control strategies\\
		
		$C_{8}$ & \cite{Wang2022AFramework}, \cite{Ma2020Cooperative}, \cite{Lee2021Design}, \cite{Sawant2021Robust}, \cite{Xing2020Compensation}, \cite{MarkVollrath2011Accident}, \cite{Tajeddin2016GMRES} & TFT requires extensive manual work, leading to high costs and time. & AI sensor optimization \\
		
		$C_{9}$ & \cite{Yang2021Research}, \cite{Zhu2020Synthesis}, \cite{Wu2019Cooperative}, \cite{VMilanes2014Modeling}, \cite{Lin2020Robust}, \cite{Gao2020Personalized}, \cite{Chu2021Self-Learning} & Incomplete real-life testing may violate regulations & Prioritize comprehensive real-life testing to ensure regulatory compliance\\
		
		$C_{10}$ & \cite{Wasserburger2020Probability}, \cite{Zhang2021Data-Driven}, \cite{Yu2022Safety}, \cite{Zhang2020ControlDesign}, \cite{Alotibi2021Anomaly}, \cite{boddupalli2022resilient}, \cite{de2021advanced} & Ignoring communication loss & develop robust communication protocols\\  
		\bottomrule
	\end{tabular} 	
\end{table*}

%% file: includes/sections/concluslions.tex
\section{Conclusions}
\label{sec:con}

In conclusion, the ACC technology helps achieve advancements in the automation of transportation. This helps in improving safety, optimizing traffic, etc. However, there are a lot of challenges that hinder the widescale adoption of Avs. Through our work, we aim to tackle these issues. In the beginning, we introduced the ACC and its different types and discussed the main challenges faced by the ACC. Later through our novel taxonomy, we derive the gaps in the domain of cruise control. Lastly, we suggest a set of future directions that will help in bringing notable advancements in the design of the ACC systems, which will help in achieving a sustainable automated transportation ecosystem.

%% file: includes/sections/acknowledgement.tex
\section*{Acknowledgments}

The authors thank anonymous reviewers for their insightful comments on improving the quality.